\pdfoutput=1
\documentclass[10pt,a4paper,ptm,twoside]{article}

\usepackage[utf8]{inputenc}
\usepackage[english]{babel}

\usepackage{amssymb}
\usepackage{amsmath}
\usepackage{amsfonts}

\usepackage{xcolor,graphicx,soul,multirow}
\usepackage[round]{natbib}
\usepackage[affil-it]{authblk}
\usepackage[outline]{contour}
\contourlength{.2pt}
\usepackage[breaklinks=true,bookmarks=true,colorlinks=true,
            linkcolor=blue,citecolor=blue,urlcolor=blue]{hyperref}
\usepackage{nomencl}
\makenomenclature

\newcommand{\figdir}{./}
\newcommand{\refdir}{./}
\newcommand{\mmc}[1]{{\color{blue!}\st{}}}
\newcommand{\mmt}[1]{{\color{black!}#1}}

\newcommand{\mmr}[2]{{\color{black!} #2}}

\newcommand{\mun}[1]{{\color{red!}\textsc{#1}}}

\newcommand{\putpic}[3]{\put(#1){\includegraphics[trim=0cm 0cm 0cm 0cm, clip, width=#2\textwidth]{\figdir #3}}}

\usepackage{geometry}
\newlength\halflineskip
\newlength\affilskip
\usepackage{fancyhdr}
\usepackage[font=small,labelfont=sc]{caption}
\usepackage{titlesec}
\titleformat{\section}
  {\bf\large}
  {\thesection}{1em}{\vspace*{-.2cm}}
\usepackage{etoolbox}
\AtBeginEnvironment{tabular}{\small}

\usepackage{array}
\newif\ifref
\reftrue
\newif\ifnotes
\notesfalse
\newif\ifall
\alltrue
\newif\ifzero
\newif\ifone
\newif\iftwo
\newif\ifthree
\newif\iffour
\newif\iffive
\newif\ifsix
\newif\ifseven
\newif\ifeight
\newif\ifnine
\newif\iften
\newif\ifeleven
\newif\iftwelve
\newif\ifthirteen
\newif\iffourteen
\newif\iffifteen
\newif\ifsixteen
\newif\ifseventeen
\newif\ifeighteen
\newif\ifnineteen
\newif\iftwenty
\newif\iftwentyone
\newif\iftwentytwo
\newif\iftwentythree
\newif\iftwentyfour
\newif\iftabs
\newif\iftimephase
\newif\ifnumericalmethod
\zerofalse
\onefalse
\twofalse
\threefalse
\fourfalse
\fivefalse
\sixfalse
\sevenfalse
\eightfalse
\ninefalse
\tenfalse
\elevenfalse
\twelvefalse
\thirteenfalse
\fourteenfalse
\fifteenfalse
\sixteenfalse
\seventeenfalse
\eighteenfalse
\nineteenfalse
\twentyfalse
\twentyonefalse
\twentytwofalse
\twentythreefalse
\twentyfourfalse
\timephasefalse
\numericalmethodfalse
\ifall
\zerotrue
\onetrue
\twotrue
\threetrue
\fourtrue
\fivetrue
\sixtrue
\seventrue
\eighttrue
\ninetrue
\tentrue
\eleventrue
\twelvetrue
\thirteentrue
\fourteentrue
\fifteentrue
\sixteentrue
\seventeentrue
\eighteentrue
\nineteentrue
\twentytrue
\twentyonetrue
\twentytwotrue
\twentythreetrue
\twentyfourtrue
\tabstrue
\timephasetrue
\numericalmethodtrue
\fi
\tabsfalse
\timephasefalse
\numericalmethodfalse
\newcommand\emt[1]{\ensuremath{#1}}
\newcommand\mob{\emt{    \psi}}           % Mobility number
\newcommand\denss{\emt{  \varrho_s}}      % Grain density
\newcommand\densf{\emt{  \varrho}}        % Fluid density
\newcommand\g{\emt{      g}}              % Gravitational acceleration
\newcommand\vs{\emt{     v_s}}            % Gravitational velocity
\newcommand\ds{\emt{     d}}              % Sphere diameter 1
\newcommand\s{\emt{      s}}              % Relative density
\newcommand\U{\emt{      U_0}}            % Free-stream velocity
\newcommand\sstr{\emt{   \widetilde{u}}}  % Steady streaming magnitude
\newcommand\fse{\emt{    \ell_f}}         % Fluid semi-excursion
\newcommand\kvis{\emt{   \nu}}            % Kinematic viscosity fluid
\newcommand\dvis{\emt{   \mu}}            % Dynamic viscosity fluid
\newcommand\del{\emt{    \delta}}         % Stokes BL thickness
\newcommand\om{\emt{     \omega}}         % Anguilar frequency
\newcommand\TT{\emt{     T}}              % Oscillation period
\newcommand\Rdel{\emt{   Re_\del}}        % Reynold \delta
\newcommand\Rd{\emt{     Re_\ds}}         % Reynold particle
\newcommand\Ga{\emt{     Ga}}             % Galileo Number
\newcommand\Kc{\emt{     K_c}}            % Keulegan-Carpenter number
\newcommand\svf{\emt{    \phi_s}}         % Solid volume fraction
\newcommand\bsur{\emt{   \eta}}           % Bed surface
\newcommand\bprP{\emt{   \bsur_\phi}}     % Bed profile - definition 1
\newcommand\bprPrms{\emt{\bsur_{rms}}}    % rms of bed profile
\newcommand\bprE{\emt{   \bsur_{\cal E}}} % Bed profile - definition 2
\newcommand\wln{\emt{    \lambda_\bsur}}  % Dominant wavelength of ripples
\newcommand\wlns{\emt{   \lambda_{\sstr_i}}} % Dominant wavelength of the steady streaming
\newcommand\pse{\emt{    \ell_s}}         % Particle semi-excursion
\newcommand\pvel{\emt{   u_{1s}}}         % Particle velocity
\newcommand\pdr{\emt{    F_{1s}}}         % Particle drag
\newcommand\prf{\emt{    F_{ref}}}        % Reference force
\newcommand\phs{\emt{    \chi}}           % Transport initiation phase for top-particles
\newcommand\tin[1]{\emt{ t_{#1}^{(in)}}}  % Start phase of particle tracking
\newcommand\SSp[1]{\emt{ S_{#1#1}}}       % Spatial Spectrum

\newcommand\sav[1]{\emt{ \left\langle #1\right\rangle}} % Space average
\newcommand\vc[2]{\emt{  #1_1^{#2},#1_2^{#2},#1_3^{#2}}}          % vector components
\newcommand\Lx{\ensuremath{L_{x_1}}}
\newcommand\Ly{\ensuremath{L_{x_2}}}
\newcommand\Lz{\ensuremath{L_{x_3}}}
\newcommand\Npt{\ensuremath{N_s}}         % No. of  parts
\newcommand\tautot{\emt{ \tau_{tot}}} % Total shear stress
\newcommand\tauf{\emt{ \tau_{f}}} % fluid shear stress
\newcommand\taup{\emt{ \tau_{p}}} % particle shear stress
\newcommand\shields{\emt{\theta}} % Shields number
\newcommand\qpmean{\emt{q_s}} % mean part. flowrate
\title{\bf Direct numerical simulations of ripples in an oscillatory flow}

\author[1]{{\bf MARCO MAZZUOLI}}
\author[2]{{\bf AMAN G. KIDANEMARIAM~\footnote{\textit{Current address:} Federal Waterways Engineering and Research Institute (BAW), Karlsruhe, Germany}}}
\author[2]{{\bf MARKUS UHLMANN}}
\affil[1]{{\small Department of Civil, Chemical and Environmental Engineering (DICCA), University of Genoa, %\\[\affilskip]
Via Montallegro 1, 16145 Genova, Italy}}
\affil[2]{{\small Institute for Hydromechanics, Karlsruhe Institute of Technology,% \\[\affilskip]
76131 Karlsruhe, Germany}}

\date{October 2018}

\begin{document}

\maketitle

\vspace{.5cm}

%\begin{frontmatter}



%\title{Direct numerical simulations of small-scale bedforms in an oscillatory flow}


%\author[A1]{Marco Mazzuoli\corref{cor1}}
%\cortext[cor1]{corresponding author}
%\ead{marco.mazzuoli@unige.it}
%\author[A2]{Aman Kidanemariam}
%\ead{aman.kidanemariam@kit.edu}
%\address[A1]{Department of Civil, Chemical and Environmental Engineering (DICCA), Via Montallegro 1, 16145, Genoa, Italy}
%\address[A2]{Institute for Hydromechanics, Karlsruhe Institute of Technology, 76131 Karlsruhe, Germany}

\begin{abstract}
\mmr{%
Sea ripples are small-scale bedforms originating from the interaction of an 
oscillatory flow with a movable bed. %
The formation of ripples is investigated by means of two direct numerical 
simulations of the incompressible Navier-Stokes equations. %
%at moderate values of the Reynolds number. %
%
The flow is fully resolved around the sediment particles which consist of rigid 
%ak
%inelastic 
spheres %
\mmt{%
of diameter comparable with the thickness of the boundary layer%
}%
. %
The simulations reproduce with fair fidelity two laboratory experiments, differing %which differed 
in the amplitude and frequency of fluid oscillations, where rolling-grain 
ripples were observed to develop. %
\mun{I would remove from here $\rightarrow$}
After an initial transient of 
%mu
%a few 
\mmr{several}{approximately ten} %
oscillation periods, patterns of particles 
%mu
%are observed 
form 
which then coarsen and give rise to ripples. %
%
%The boundary layer does not separate on the lie side of ripples.
%
%In particular, ripples form more rapidly in the simulation characterised by 
%the larger Reynolds number. %
%
%In the latter case the growth of bed surface disturbances follows an 
%exponential trend. %
%
The evolution of the bed surface is investigated and 
%mm
%surficial 
the 
sediment particles at the flow-bed interface 
are tracked %
\mmr{%
in order to be able to %
}{%
in order to %
}%
describe their dynamics through the wave cycles. %
\mun{$\leftarrow$ to here. }
\mmr{%
The picture that arises confirms the theoretical framework underlying 
the mechanism of selection of ripple wavelength. %
}{%
The results confirm that ripple wavelength is determined by the action of a 
steady streaming consisting of space-periodic recirculating cells which tend 
to accrete ripple crests and scour ripple troughs. %
The growth rate of ripples increases if the maximum bed shear stress is about 
the critical condition to incipient sediment transport, sediments being more sensitive 
to the effect of the steady streaming. %
}%
%
%\mmr{}{%
The knowledge on the origin of ripples is presently enriched by insights
and measurements of quantities difficult to obtain %
%mm
%precisely 
in the laboratory, %
\mmt{%
like particle forces, the statistics of particle motion and the bed shear stress. %
}%
%}%
%
Finally, 
\mmc{the sediment flow rate is computed and related to the bed shear stress and 
to the pressure gradient. }%
\mmt{%
it is found that, for sand seafloor and typical wave periods characterising 
morphogenetic wind waves, in the absence of turbulence viscous effects
are predominant and the sediment flow rate is affected both by the bed 
shear stress and by the wave-induced pressure gradient. %
}%
%
%Beyond the bed shear stress, also the pressure gradient is found to play a significant 
%role.
%
}{%
Sea ripples are small-scale bedforms which originate from the
interaction of an oscillatory flow with an erodible sand bed.  The
phenomenon of sea ripple formation is investigated by means of direct
numerical simulation in which the sediment bed is represented by a
large number of fully-resolved spherical grains (i.e, the flow around
each individual particle is accounted for).  Two sets of parameter
values (differing in the amplitude and frequency of fluid
oscillations, among other quantities) are adopted which are motivated
by laboratory experiments on the formation of laminar rolling-grain
ripples. %which exhibit no flow separation.
The knowledge on the origin of ripples is presently enriched by
insights and by providing fluid- and sediment-related quantities 
that are difficult to obtain in the laboratory (e.g. particle forces, 
statistics of particle motion, bed shear stress). %
In particular, %
detailed analysis of flow and sediment bed evolution has confirmed
that ripple wavelength is determined by the action of steady
recirculating cells which tend to accumulate sediment grains into
ripple crests. The ripple amplitude is observed to
grow exponentially consistent with established linear
stability analysis theories. %
Particles at the bed surface %, which have been tracked, 
exhibit two kinds 
of motion depending on their position with respect to the 
recirculating cells: particles at ripple crests are significantly 
faster and show larger excursions than those lying on ripple troughs. %
In analogy with %the phenomenon of segregation of 
%sediments of different size or density, 
segregation phenomenon of polydisperse sediments %
the non-uniform distribution of the velocity field 
promotes the formation of ripples. %
The wider the gap between the excursion of fast and slow particles, the
the larger the resulting growth rate of ripples. %
%larger the growth rate of ripples results. %
%
Finally, it is revealed that, in the absence of turbulence, the
sediment flow rate is driven by both the bed shear stress and the
wave-induced pressure gradient, the dominance of each depending on the
phase of the oscillation period. In phases of maximum bed shear
stress, the sediment flow rate correlates more with the Shields number
while the pressure gradient tends to drive sediment bed motion during
phases of minimum bed shear stress.
}%

%------------------------------------------
%* by 'phase-reduced' I mean either phase-average, phase-min phase-max. I couldn't think of a better phrase.

%* did you intentionally leave out the results of section 3.2 from the abstract or do we need couple of sentences to mention the particle excursion?}
%
\end{abstract}

%\begin{keywords}



%\end{keywords}

%\end{frontmatter}

%\mbox{}
%\newpage
%\input{nomenclature.tex}
%\printnomenclature[1in]


\section{Introduction}\label{intr}

Sedimentary patterns in maritime environments are typically 
%mu 
%originated 
caused 
by different morphogenetic phenomena and 
can exhibit a wide range of spatial scales 
varying from a few centimetres to hundreds of meters. %
The capability to predict the evolution of large-scale bedforms 
is necessary, for example, to guarantee the durability of 
\mmr{%
artificial structures %
}{%
marine structures and infrastructures %
}%
as well as the equilibrium of sensitive benthic ecosystems, 
thereby preventing extraordinary catastrophic events. %
Nonetheless, the contextual presence of smaller bedforms cannot be 
neglected since morphogenetic processes occurring at different scales 
are not reciprocally independent. %
\mmc{Indeed,} It is well known that small-scale bedforms, like ripples, 
modify the structure of the flow in the vicinity of the 
%mu 
%bottom 
bed and 
they 
can significantly enhance the transport of sediments and contaminants 
near the bed \citep[e.g.][]{thibodeaux1987}. %
%
%\mmr{}{%
It has also been %
\mmr{%
recognised %
}{%
shown %
}%
that model predictions of the sediment flux 
due to the flow induced by wind waves on a plane bed 
%associated with the turbulent oscillatory flow 
can be affected by 
errors that easily exceed $100\%$ of the actual measurements 
because turbulence diffusion models currently available are not able to describe 
the turbulent convective events which characterise an oscillatory flow 
during the flow reversal \citep{Davies1997}. %
%
%Then the picture is %even poorer estimates can be obtained 
%even more complicated in the presence of ripples. %
Such discrepancies are enhanced 
\mmr{%
in %
}{%
by %
}%
the presence of ripples which can 
significantly amplify the amount of sediment set into suspension. %
%}%
%

\mmr{%
The formation of ripples is associated with the interaction of 
the flow induced by wind waves with a movable bed. %
}{%
Sea ripples originate from the action of the flow induced by wind waves 
on a movable bed under certain flow and sediment conditions. %
}%
For the sake of simplicity, let us consider the case of monochromatic wind 
waves developing over a plane bed of cohesionless sediments. %
Assuming that the linear Stokes wave theory can be used to approximate 
the irrotational flow far from the bottom, 
close to the bed the flow turns, at the leading order of approximation, 
into the oscillatory boundary layer (OBL) generated by harmonic 
oscillations of %uniform 
pressure gradient. %
\mmt{%
In real ocean, additional streaming (boundary layer streaming) 
has its origin in the existence of 
vertical velocities close to the bed originated by the non-uniformity of the flow 
beneath free-surface waves. %
Such streaming, which may also play a role on morphogenetic processes, is presently not considered. %
}%
Mathematically, the flow can be described by the incompressible
Naver-Stokes equations defined in a domain bounded by the bed surface. %
%
%\mmr{}{%
If the bed is fixed, the hydrodynamic problem is globally stable 
for moderate values of the Reynolds number as long as the fluctuations 
generated by the bed roughness do not amplify and turbulence appears. %
%}%
%
However, the material of coastal shelves often consist of
cohesionless fine- and medium-sand %(diameter $0.125\div0.5$~mm) 
%, for the typical characteristics of wind-waves in the vicinity of the coast, 
which can be easily set into motion by waves even for relatively small 
values of the Reynolds number. %
For a laminar OBL, 
%mu
%by imposing the continuity of the bed material, 
%the position of the sediment-fluid interface can 
%be determined analytically. %
when treating the sediment as a continuum, the stability problem 
can be tackled analytically. %
The resulting problem is globally unstable, %, independently 
%of the value of the Reynolds number, therefore it is expected that 
thus we can expect that 
the amplitude of a small perturbation of the bed surface grows
%the bed surface will no longer remain flat 
as the critical condition of sediment motion is reached. %and the bed surface is no longer 
%remain 
%flat. %
%

\mmc{Indeed,} \mmt{See ripples} are caused by the instability of the bed surface under the 
action of flow oscillations and consist %
%\mmr{%
in %
%}{%
%of %
%}% 
a two-dimensional waviness 
of the bed surface, the third dimension being orthogonal to flow oscillations, 
with wavelength ranging from a few (rolling-grain ripples) to some tens of 
centimetres (vortex ripples), even though three-dimensional patterns 
\mmr{were}{have} %
also %
\mmt{been} %
observed \citep[e.g. brick-pattern ripples,][]{vittori1992,pedocchigarcia2009}. %
The mechanism underlying the formation of a bottom waviness in a laminar 
OBL over a cohesionless plane bed %
\mmr{is almost}{has been fairly well} %
understood since \citet{Sleath1976a}  
observed that the interactions of a small bottom waviness (of infinitesimal 
amplitude) with the oscillatory flow induces a secondary steady flow, i.e. a 
steady streaming superimposed on the principal flow oscillations, consisting %
\mmr{in}{of} %
two-dimensional recirculating cells. %
If the steady streaming is strong enough to affect the motion of 
sediment particles, sediments tend to pile up where the streamlines 
of adjacent recirculating cells converge and to be eroded elsewhere. %
The mechanism of accumulation of sediments is balanced by the effect of 
gravitational acceleration which opposes the accretion of the waviness amplitude. %
Rolling-grain ripples are %typical bedforms originating 
the bedfroms that can form 
in a laminar OBL and 
their emergence is the first indicator that a plane bed configuration is 
evolving into rippled geometry. %
Experimentally, it was observed that, since their first appearance, ripples undergo a 
coarsening process that can stop if a stable configuration is attained, when the 
effect of gravity on sediment particles counteracts that of the steady streaming, before 
the %
\mmr{steepness of ripples}{ripple steepness, defined as the ratio between ripple height and wavelength,} %
causes flow separation \citep{stegner1999,rousseaux2004}. %
As the slope of rolling-grain ripples 
\mmr{is}{becomes} %
large enough to %promote 
trigger %
the %
separation of the flow from their crests, vortex ripples form which 
are characterized by steeper slopes and larger 
\mmr{%
amplitude %
}{%
height and wavelength %
}%
than rolling-grain ripples. %

As long as the %the flow is laminar, 
boundary layer does not separate from the bed surface, 
\mmr{%
the problem of determining the growth rate of wavy perturbations of the bottom surface
can be tackled by means of linear stability analysis. %
}{%
the growth rate of wavy bedforms may be determined through linear stability analysis. %
}%
%of the hydrodynamic problem at the flow-bed interface can be performed to predict the wavelength first appear from a plane bed. %
%
This approach was first adopted by \citet{lyne1971} and \citet{Sleath1976a} 
under the hypothesis of large fluid displacement oscillations, i.e. much larger 
than the wavelength of the %
\mmr{waviness}{bedforms}, %
which however is not suitable for the case of ripples. %
Then, \citet{blondeaux1990} solved the analytic problem for %
\mmr{%
the relevant case of arbitrary values of the ratio between the fluid excursion %
%oscillations 
and the wavelength of the waviness, 
}{%
arbitrary ratios of the orbital excursion to the ripple wavelength %
}%
while \citet{vittori1990} 
extended the formulation of \citet{blondeaux1990} to the case of finite amplitude 
ripples by means of weakly-nonlinear stability analysis. %
%
%Unfortunately, l
Laboratory experiments \citep[e.g.][]{blondeaux1988} show that 
stable rolling-grain ripples can be observed only for a relatively small range 
of values of the Stokes and particle Reynolds numbers and of the mobility number 
defined, respectively, by: %
\begin{equation}
\begin{array}{lllllll}
\displaystyle \Rdel
=
\frac{\U^*\del^*}{\kvis^*} 
&,& \quad
\displaystyle \Rd
=
\frac{\U^* \ds^*}{\kvis^*} 
& \mathrm{and} & 
\displaystyle \mob
=
\frac{\U^{*2}}{\vs^{*\,2}} 
%
%\s=\frac{\denss^*}{\densf^*} 
\:\:,
\end{array}
\label{eq1}
\end{equation} 
where $\U^*$ denotes the amplitude of free-stream velocity oscillations, 
$\del^*=\sqrt{2\kvis^*/\om^*}$ denotes the conventional thickness of a viscous 
oscillatory boundary layer \citep{sleath1984} and $\om^*$ the angular frequency of 
flow oscillations. %
The quantity $\vs^*$ is often referred to as \textit{gravitational velocity} of sediment particles 
and is defined as 
\begin{equation}
\vs^* 
= 
\sqrt{\left(\frac{\denss^*}{\densf^*}
-
1 \right) \g^* \ds^*}
\label{eq:vs}
\end{equation}
where $\g^*$ indicates the modulus of gravitational acceleration, $\denss^*$ 
and $\ds^*$ the density and the nominal diameter of sediment grains 
while $\densf^*$ and $\kvis^*$ are the density and the kinematic viscosity 
of the fluid. %
The period of the flow oscillations is denoted by $\TT^*$ and equal to $\pi/\om^*$.
The star superscript is used to denote dimensional quantities and distinguish them from 
dimensionless ones. %
The parameters~\eqref{eq1}, along with the specific gravity $\s=\denss^*/\densf^*$, 
can be chosen to determine the parameter space %
\mmr{of the problem of}{for} %
sediment transport %
\mmr{for}{with} %
spherical particles in %
\mmt{the} %
absence of bedforms. %
Alternatively, the Galilei number $\Ga$ is often used in the particulate flow and suspension communities, which is related to $\mob$ and $\Rd$ through the expression $\Ga=\Rd/\sqrt{\mob}$, as well as the 
Keulegan-Carpenter number, $\Kc=\Rdel^2/(2\,\Rd)$, that is defined as the ratio between the semi-excursion of the fluid far from the bed, $\fse^*=\U^*/\om^*$, and the diameter of sediment particles. %
As the \emph{average ripple steepness} %
\mmr{, of the ripples}{}%
\mmc{defined as the ratio between 
the height and the wavelength,} %
exceeds 
\mmr{approximately the value}{the threshold} %
$0.1$ identified empirically by \citet{sleath1984}, 
the flow separates from %
\mmr{their}{the ripple} crests and computations of the %
\mmr{evolution of ripples}{ripple evolution} %
can %
\mmr{be made only by numerical means. }{only be made numerically}. %
For instance, \citet{scandura2000} studied numerically the interaction of an 
oscillatory flow with a wavy wall, characterized by steepness $\sim 0.1$, for 
values of $\Rdel$ ranging between $42$ and $89$, and observed the flow separation 
from the crests of the wall and the appearance of three-dimensional %
\mmr{%
disturbances %
}{%
vortex structures. %
}%
However, \citet{scandura2000} concluded that %the evolution of the 
movable %
bed should be considered in the simulations to obtain results relevant for %
the problem of sediment transport. %

Since the evolution of the bed surface is not known a priori but results from the
coupling between the fluid and sediment dynamics, a discrete approach seems 
more suitable to investigate the mechanics of sediment particles in an OBL. %
%
%Indeed, in the reality, at the early stages of the formation of ripples, 
%%the surficial sediment particles characterise the small perturmations of the plane bed and their 
%the growth of perturbations associated with bed-surface irregularities seems to be 
%initially driven by the dynamics of a small amount of sediments picked up from 
%the bed surface, which barely fulfil the continuum hypothesis (e.g. how bed surface 
%and sediment velocity are defined?). %
%
In order to investigate the origin of ripples and test the capability of the numerical 
approach to catch the basic physics of the sediment transport,
\citet{mazzuoli2016a} performed Direct Numerical Simulations (DNSs) of an oscillatory 
flow both over smooth and rough %
\mmr{wall}{walls} %
with movable spherical beads on top of it. %
The values of the parameters were chosen similar to those of laboratory experiments 
where the formation either of sediment patterns \citep{hwang2008} or of rolling-grain 
ripples \citep{blondeaux1988} had been observed. %(in the laminar flow regime). %
\citet{mazzuoli2016a} 
considered %spherical particles 
identical beads initially aligned along the direction of flow oscillations and 
observed that, %
\mmr{in a few}{within a few} %
oscillation periods, they %aligned 
rearranged in chains orthogonal to the flow oscillations, equispaced 
%in the direction orthogonal to flow oscillations originating chains almost equispaced 
by a distance comparable to 
that measured in the experiments. %
%the wavelength of the ripples measured in the experiments. %
%
%Furthermore, q
Qualitatively, the mechanism of formation of the chains was %
\mmr{%
found %almost independent of 
barely affected by the number of beads 
%(as long as their solid volume fraction at the bottom level was much smaller than $1$) 
and eventually by the presence of bottom roughness 
consisting in %
}{%
not very sensitive to the number of beads or the presence of the bottom roughness consisting of %
}%
beads closely packed and fixed on the 
\mmr{%
wall. %
}{%
bottom. %
}%
%characterised by the same size as the beads. %
%
%In fact, s
Steady recirculating cells of different sizes initially developed, %in all the cases, %but, 
but only recirculating cells compatible with the wavelength of the chains of beads 
were promoted and could be observed at the final stages of the simulations. %
%
%This suggests that the dynamics of the fluid and the solid phases are strictly 
%connected and cannot be uncoupled, whereas additional ingredients like particle-particle 
%collisions, may only result in quantitative (e.g. lag) effects. %
%

%The dynamics of bead-chains 
%In principle, t
Since the process of formation of chains of spheres 
is basically different from that 
of ripples, %
\mmt{due to }%
gravity playing different roles in the two cases, %
%Therefore, 
two of the experiments of \citet{blondeaux1988}, where rolling-grain ripples formed, 
were reproduced by means of DNS and are presently described. %
\mmt{%
The values of the relevant dimensional parameters characterising the experiments are reported in table~\ref{tab4}. %
}%
%%
%*** Table 4 *******************************************************
\begin{table}[ht]
	\begin{center}
		\begin{tabular}{l c c c c c c}
		\hline
		\multirow{1}{*}{} & 
		\multirow{1}{*}{$\TT^*\ [$s$]$} & 
		\multirow{1}{*}{$\U^*\ [$m/s$]$} & 
		\multirow{1}{*}{$\fse^*\ [$m$]$} & 
		\multirow{1}{*}{$\del^*\ [$mm$]$} & 
		\multirow{1}{*}{$\ds^*\ [$mm$]$} & 
		\multirow{1}{*}{$\denss^*/\densf^*$} \\ 
		\hline
		run 1 & $1.86$ & $0.16$ & $0.8$ & $0.05$ & $0.2\pm0.06$ & $2.65$ \\
		run 2 & $0.95$ & $0.13$ & $0.6$ & $0.02$ & $0.2\pm0.06$ & $2.65$ \\
		\hline
		\end{tabular}
	\end{center}
	\caption{%
	\mmt{
	Parameters for \citet{blondeaux1988}'s experiments presently considered. %
	From left to right: the oscillation period, the amplitude of free-stream 
	velocity oscillations, the stroke or fluid semi-excursion, the thickness 
	of the Stokes boundary layer, the particle diameter 
	%($0.06$ is the standard deviation of the grain distribution) 
	and the particle specific gravity. %
	The kinematic viscosity of the fluid was approximately equal to $10^{-6}~$m$^2/$s. %
	}%
	}%
	\label{tab4}
\end{table}
%
%\textbf{%
  In particular, the present investigation is aimed at: 
  (i) showing that laboratory experiments of the formation of ripples can be reproduced by DNS, 
  (ii) obtaining accurate values of quantities that are difficult to be measured in the laboratory %
\mmt{%
(e.g. particle forces and trajectories, steady streaming intensity, bed shear stress, the sediment flow rate), %
}%
  (iii) investigating the dynamics of sediment particles, and 
  (iv) relating the sediment transport to mean flow quantities. 
%}
%%%%%%%%%      FINO QUI     %%%%%%%%%%%%

%
In the following, the numerical method is briefly described while the results are discussed in \S\ref{resl}. 
Finally, conclusive remarks are drawn in \S\ref{conc}. % to close the present contribution. %
\section{Formulation of the problem and numerical approach}\label{appr}

The OBL (over a smooth wall) can be generated in the laboratory %as an effect of 
by the harmonic motion of a piston which produces a uniform pressure gradient 
through the fluid, in a duct with sufficient depth and breadth to prevent 
undesirable boundary effects. %
%make negligible the mutual effects of the boundaries. 
%
Typically the axis of the duct develops along a U-shape profile in order to exploit 
the support of gravity, while %and support with the piston the natural oscillations 
%with period proportional to the square root of the duct length. %
%
only the flow field in the central section of the U-tube, in the vicinity of the bottom, 
%side plate 
is investigated. %
The time-development of the pressure gradient driving the flow is described by
\begin{equation}
\label{pres-g}
\frac{\partial p_f^*}{\partial x^*_1} = - \varrho^* U^*_0 \omega^*
\sin (\omega^*t^*); \ \ \ \ \frac{\partial p_f^*}{\partial x^*_2} =
0; \ \ \ \ \frac{\partial p_f^*}{\partial x^*_3} = 0
\end{equation}
where $t^*$ is the time variable and $(\vc{x}{*})$ is a Cartesian coordinate 
system with origin at the %
\mmr{%
wall %
}{%
bottom of the domain, %
}%
the $x_1^*$-axis parallel to the flow oscillations 
and the $x_2^*$-axis pointing the upward wall-normal %
\mmt{%
(i.e. bottom-normal) %
}%
direction. %
The total pressure can be expressed by the sum:
\begin{equation}
p^*_{tot}(\vc{x}{*},t^*)
\ =\ 
\dfrac{Re_\delta}{2}p^*_f(t^*)
\ +\  
%\widetilde{p}^*
%+ 
p^*(\vc{x}{*},t^*)\: ,
\label{pres-tot}
\end{equation}
where $p_f^*$ is equal to the right hand side of the first component of \eqref{pres-g} multiplied by $x_1^*$ and $p^*$ denotes the pressure 
%mu
%due to the boundary conditions. %in the OBL. 
in the boundary layer. %
%
%\mmr{%}{%
Then, $p^*$ (as well as any other flow quantity) can be further split into the sum of two contributions: %
\begin{equation}
p^*(\vc{x}{*},t^*)
\ =\ 
\overline{p}^* 
\ +\  
p'^* (\vc{x}{*},t^*)\: ,
\label{pres-p}
\end{equation}
the flat overbar indicating the statistical average operator (the ensemble average or the phase average, i.e. the average computed at corresponding phases of the oscillation period, if the flow and bed evolution are at the equilibrium) and $p'^*$ the corresponding fluctuating part. %
%}%
Let the %
\mmr{%
wall %
}{%
bottom %
}% 
(i.e. the plane $x_2^*=0$) be equipped with a bed of monosized spherical heavy particles of diameter $d^*$ initially arranged in multiple superimposed plane layers. %
The dynamics of the particles is dictated by the collective influence of gravity, 
collision and hydrodynamic forces. %
Hydrodynamic force, in turn, results from the combination of pressure and viscous contributions. %
The pressure gradient \eqref{pres-g} drives both the motion of the fluid and of the solid particles while the fluctuations of pressure, denoted by $p'^*$ in equation \ref{pres-p}, can be associated both with turbulence and with the motion of particles. %
%
%\mmr{}{%
Since the ensemble average is not feasible with a single simulation while the ``equilibrium state'' is presently never attained, different spatial-average operators are adopted to estimate the average quantities. %
%It is also useful to 
\mmr{Let us introduce %here also the spatial-average 
t}{T}he operator 
%}%
$\langle\cdot\rangle_{\alpha}^{(i)}$ %
\mmr{which}{} %
denotes the average of the 
argument 
%ak
%function 
performed along the direction $\alpha\equiv x_1,\,x_2$ or $x_3$, 
%($x_\alpha$-direction average), 
or along two directions, e.g. $\alpha\equiv x_1x_3$ indicates 
the horizontal plane (\emph{plane average}), or over a three-dimensional sub-space 
$\alpha\equiv{\cal V}$ (\emph{volume average}). For the sake of simplicity, omitting 
$\alpha$ implicitly indicates that the plane average is performed. %
The superscript $(i)$, if present, indicates that the flow field has 
been split into a number of bins either along the streamwise direction, 
equispaced by $h_1^*=2d^*$, or along the wall-normal direction, equispaced 
by $h_2^*=d^*$, and that the average is computed over the $i$-th bin. %
A similar notation is adopted for particle-related quantities to indicate 
the average over a set of particles ($\alpha\equiv s$) or the time-average over 
each half-period ($\alpha\equiv \TT/2$). %

On the basis of purely dimensional considerations, for the present flow configuration, 
a generic hydrodynamic quantity $\mathcal{F}^*$ can be expressed as a function 
$\mathcal{F}^*(x_i^*,t^*;\om^*,\U^*,\ds^*,\g^*,\\ \dvis^*,\densf^*,\denss^*)$, $i=1,\,2,\,3$, 
where $\dvis^*=\densf^*\kvis^*$ denotes the dynamic viscosity of the fluid. %
%can be expressed as a function of the dimensionless quantities $(x_i,t;\Rdel,\Rd,\mob,\s)$,  where $\dvis^*$ denotes the dynamic viscosity of the fluid and $i=1,\,2,\,3$. %
%By using $\om^*$, $\densf^*$ and $\dvis^*$, t
%The number of independent arguments can be reduced by using $\om^*$, $\densf^*$ and $\dvis^*$ 
The present choice is to use $\om^*$, $\densf^*$ and $\dvis^*$ to reduce the number of dimensionally dependent arguments and obtain 
%which yields 
the corresponding dimensionless quantity $\mathcal{F}(x_i,t;\Rdel,\Rd,\mob,\s)$ which depends on 
the numbers introduced in \S\ref{intr}. %
The values of the numbers $\Rdel,\Rd,\mob$ and $\s$ for the present simulations are indicated in table~\ref{tab1}. %
\mmt{%
Note that the specific gravity between the runs differs by $7\%$ which is not
expected to play a significant role. %
}%
%
%*** Table 1 *******************************************************
\begin{table}%[ht]
	\begin{center}
		\begin{tabular}{l c c c c c c}
		\hline
		\multirow{1}{*}{} & 
		\multirow{1}{*}{$Re_\delta$} &
		\multirow{1}{*}{$Re_d$} & 
		\multirow{1}{*}{$\mob$} & 
		\multirow{1}{*}{$Ga$} & 
		\multirow{1}{*}{$\Kc$} & 
		\multirow{1}{*}{$s$} \\ 
		\hline
		run 1 & $128$  & $33.3$ & $9.69$ & $10.7$ & $246$ & $2.46$ \\
		run 2 & $71.7$ & $17.4$ & $7.88$ & $6.2$  & $148$ & $2.65$ \\
		\hline
		\end{tabular}
	\end{center}
	\caption{Summary of flow parameters for the present runs.}
	\label{tab1}
\end{table}
%*** Table 2 *******************************************************
\begin{table}
	\begin{center}
		\begin{tabular}{l c c c c c c c c c}
		\hline
		\multirow{1}{*}{run} & 
		\multirow{1}{*}{$\Lx$} & 
		\multirow{1}{*}{$\Ly$} &
		\multirow{1}{*}{$\Lz$} &
		\multirow{1}{*}{$n_{x_1}$} & 
		\multirow{1}{*}{$n_{x_2}$} &
		\multirow{1}{*}{$n_{x_3}$} &
%		\multirow{1}{*}{$\Delta x$} &
		\multirow{1}{1.3cm}{\centering{$\Delta t$}} &
		\multirow{1}{*}{$t_{fin}$} &
		\multirow{1}{1.3cm}{\centering{$\Npt$}} \\ 
		\hline
		run 1 & $53.2$ & $26.6$ & $26.6$ & $2048$ & $1024$ & $1024$ & $1.96\cdot 10^{-4}$ & $41\pi$ & $257138$ \\
		run 2 & $49.1$ & $18.4$ & $24.5$ & $2048$ & $768$  & $1024$ & $2.62\cdot 10^{-4}$ & $58\pi$ & $223442$ \\
		\hline
		\end{tabular}
	\end{center}
	\caption{Domain and time discretisation for the present runs. The final time of the simulations is denoted by $t_{fin}$ while $\Npt$ is the number of spheres used in each run.} 
	\label{tab2}
\end{table}
%*******************************************************************
%
%Similarly, 
Thus, the incompressible Navier-Stokes equations can be expressed in a dimensionless form by introducing the following variables: 

%\begin{equation}
\begin{align}%{lllll}\smallskip
(\vc{x}{}) = \frac{(\vc{x}{*})}{\delta^*} \ &;\ \ \ t = t^* \omega^*\ \ ;  & \label{var}\\
(\vc{u}{}) = \frac{(\vc{u}{*})}{U^*_0} 
\ &;\ \ \ 
p = \frac{p^*}{\varrho^*(U^*_0)^2} 
\ \ ;& \ 
(\vc{f}{}) = \frac{(\vc{f}{*})}{U^*_0\omega^*} 
\ .\nonumber
\end{align}
%\end{equation}
In (\ref{var}), %$t^* $ is time, 
$\vc{u}{*}$ are the fluid velocity
components along the $x^*_1$-, $x^*_2$- and $x^*_3$-directions, respectively, and $\vc{f}{*}$ are the components of the body force. %, while $\delta^*$, $\omega^*$ and $U_0$ denote the conventional thickness of the viscous boundary layer, the frequency of flow oscillations and the maximum free-stream velocity, respectively (see definition provided in \S\ref{intr}).
Hence, the dimensionless continuity and Navier-Stokes equations read: %
%hydrodynamic problem reads: %
\begin{align}
&\frac{\partial u_j }{\partial x_j}
=
0 \label{equ2}\\
&\frac{\partial u_i}{\partial t} 
+ 
\frac{Re_{\delta}}{2} u_j \frac{\partial u_i}{\partial x_j} 
= 
- \frac{Re_{ \delta}}{2} \frac{\partial p}{\partial x_i} 
+ \delta_{i1} \sin(t) 
+ 
\frac{1}{2} \frac{\partial^2 u_i }{\partial x_k \partial x_k} 
+ 
f_i\label{equ1}
\end{align}
where Einstein's convention on the summation is used.
%%
%%
%%    Figure 21
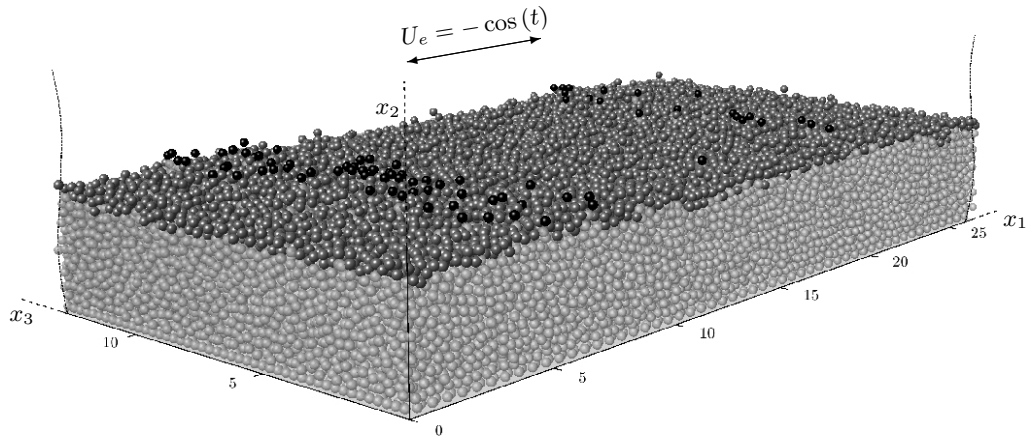
\begin{figure}
\begin{picture}(0,170)(0,0)
  \iftwentyone
  \put(0,-7){
    \put(0,0){\includegraphics[trim=0cm 0cm 0cm 0cm, clip, width=1\textwidth]{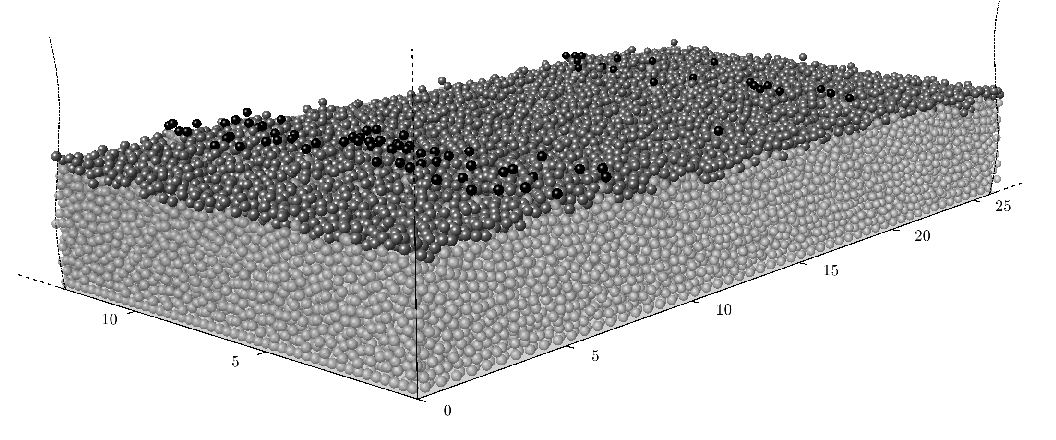}}
  \put(2,52){\small $x_3$}
  \put(374,88){\small $x_1$}
  \put(139,130){\small $x_2$}
  \put(60,20){
  \put(90,130){\rotatebox{10}{\vector(1,0){50}}}
  \put(140,138.5){\rotatebox{10}{\vector(-1,0){50}}}
  \put(88,136){\rotatebox{9}{\small $U_e=-\cos{(t)}$}}
  }
  }
  \fi
\end{picture}
\caption{
Sketch of a simulation (detail of the computational domain). Different colours are used to distinguish \textit{top-layer particles} (dark grey) from \textit{crest particles} (black). Most of the particles shadowed by light grey %are resting throughout the simulation or 
exhibit negligible displacements with respect to top-layer particles throughout the simulation. %
}
\label{fig21}
\end{figure}
It can be noted that the Reynolds number $\Rdel$ is the only dimensionless parameter which 
is based upon purely hydrodynamic quantities and controls the momentum equation~\eqref{equ1}, 
while sediments enter the problem through the boundary conditions. %
Concerning the simulation of the particle motion, the main dimensionless control parameters are 
the specific gravity of sediments $s$, the sphere Reynolds number $Re_d$ and the mobility number $\mob$ that are defined in \S\ref{intr}. %
%
%Also the Galilei number $\Ga$ is often used in the particulate flow and suspension communities, which is related to $\mob$ and $\Rd$ through the expression $\Ga=\Rd/\sqrt{\mob}$ (cf. table~\ref{tab1}). %
%

%
The domain where equations \eqref{equ2} and \eqref{equ1} are solved numerically 
is a cuboid space of dimensions $\Lx$, $\Ly$ and $\Lz$ in the streamwise, 
wall-normal and spanwise directions, respectively, which are indicated in table~\ref{tab2}. %
While periodic conditions are applied at the boundaries in the streamwise and spanwise 
directions, the no-slip condition is forced at the %
\mmr{%
wall %
}{%
bottom, %
}%
viz. 
\begin{equation}
\label{bc1}
(u_1,u_2,u_3)= (0,0,0) \ \ \ \mbox{at} \ \ \ x_2 = 0
\end{equation}
and the free slip condition is forced at the upper boundary: 
\begin{equation}
\label{bc1x}
\left( \frac{\partial u_1}{\partial x_2},\frac{\partial u_3}{\partial x_2} \right)= (0,0) ; \ \ \ \  \ \ \ u_2=0 \ \ \ \ \ \  \mbox{at} \ \ \ x_2 = \Ly\: .
\end{equation}
The dimension $\Ly$ of the domain is chosen large enough to guarantee a vanishing 
shear stress far from the bottom. %
The choice of the streamwise and spanwise dimensions of the computational domain, 
$\Lx$ and $\Lz$, can significantly affect the process of formation of the bedforms. %
In particular, as recently pointed out by \citet{kidanemariam2017}, the choice of $\Lx$ allows the development 
of bedforms characterised by wavelengths equal to 
$\left(1,\,\frac{1}{2},\,\frac{1}{3},\,\frac{1}{4},\,\ldots\right)\Lx$, thus the evolution 
of the geometrical properties of bedforms are expected to be markedly discontinuous with respect to time. %
For instance, the larger $\Lx$ the smoother %the increase %or shortening of the bedform wavelength appears. %
the evolution of bedforms appears. %
Therefore, the value of $\Lx$ is chosen 
as large as two times the wavelength of ripples observed in the experiments of 
%large enough to contain two ripples observed by 
\citet{blondeaux1990} 
%at the end of the respective experiments. %
when the ``equilibrium state'' was reached. %
Also the value of $\Lz$ is chosen to allow for possible formation of three-dimensional patterns, 
which were however absent in the experiments. %
A sketch of the simulations is shown in figure~\ref{fig21}. %

The hydrodynamic problem is solved throughout the whole computational domain including 
the space occupied by the solid particles. %
Indeed, the no-slip boundary condition 
at the surface of the spheres is forced by means of the (Eulerian) volume force $\vc{f}{}$ 
which is simply added to the right hand side of momentum equation~\eqref{equ1} %
\mmr{%
(immersed boundary approach). %
}{%
via the immersed boundary approach. %
}%
%
%\mmr{}{%
The flow solver consists %
\mmr{in}{of} %
the semi-implicit second-order fractional-step method, 
based on the finite difference approximation of time- and space-derivatives, 
as proposed by \citet{uhlmann2005}. %
%
%\mmr{}{%
The domain is discretised by a uniform equispaced grid of spacing $\Delta x_i^*=\ds^*/10$ in the $i$-th direction ($i=1,\,2,\,3$). %
%}%
%
The dynamics of the fluid and solid phases are coupled through the immersed boundary method 
while collision forces are computed with a soft-sphere Discrete Element Model (DEM) 
based upon a linear mass-spring-damper system. %
%
%}% \mmr
%
A detailed description of the collision model and of the validation can be found in \citet{Aman2014b}. %
The code has been recently used for different investigations by \citet{kidanemariam2014}, \citet{uhlmann2016a} and \citet{mazzuoli2017a} and, in a context similar to the present one, by \citet{mazzuoli2016a}. %

The start-up bed configuration was obtained by settling approximately $15$ layers of spheres 
(the number of spheres, $\Npt$, used for each run is indicated in table~\ref{tab2}) 
on a 
\mmr{%
smooth wall %
}{%
flat smooth bottom %
}%
while the fluid was at rest. %
One layer of spheres was preliminarily fixed on the %
\mmr{%
wall %
}{%
bottom %
}%
with a hexagonal arrangement in order to prevent 
\mmr{%
the %solid 
sliding as a block of the whole bed on the % 
}{%
the whole bed from sliding as a block along the %
}%
\mmr{%
wall %
}{%
bottom, %
}%
\mmr{that}{which} was never observed in the laboratory experiments. %
This expedient did not affect the results of the simulations because 
the particle velocity rapidly vanishes beneath the surficial layers of particles. %
%only a thin layer of spheres was really set into motion by the oscillatory flow. 
%
The spheres whose centers are located above a distance of $15~d^*$ from the %
\mmr{%
wall %
}{%
bottom %
}% 
were removed in order to obtain a flat bed surface. %

For the first wave-period of each simulation all the particles were kept fixed in order to let the 
interstitial flow develop. %during the initial transient.
%
%While the height of the computational domain should not significantly affect the dynamics of the system, the value 
Simulations~$1$ and $2$ were run for $41$ and $58$ half-cycles, respectively.
\mmt{%
Hereinafter, ``simulation'' and ``case'' can be sometimes used interchangeably in place of ``run'' referring to runs~$1$ and $2$. %
}%
%
%mu
%In the following section the obtained results are presented and discussed. %
%%

%\mmr{}{%
%Henceforth, t
The quantities closely related to the hydrodynamic problem are normalised 
as in \eqref{var}, while those more relevant for the evolution of the bed, which are directly affected 
by the particle dynamics, are preferably shown in terms of particle-related reference quantities (i.e. $\ds^*$ and $\vs^*$). %
Actually, the values of $\ds^*/\del^*$ for the two simulations are similar ($0.26$ and $0.24$ 
for runs~$1$ and $2$, respectively), thus the choice of $\ds^*$ or $\del^*$ as reference length 
scales is not practically relevant in the present configuration. %
%not relevant for the following discussion. %
%}

%%

%%
\section{Results}\label{resl}

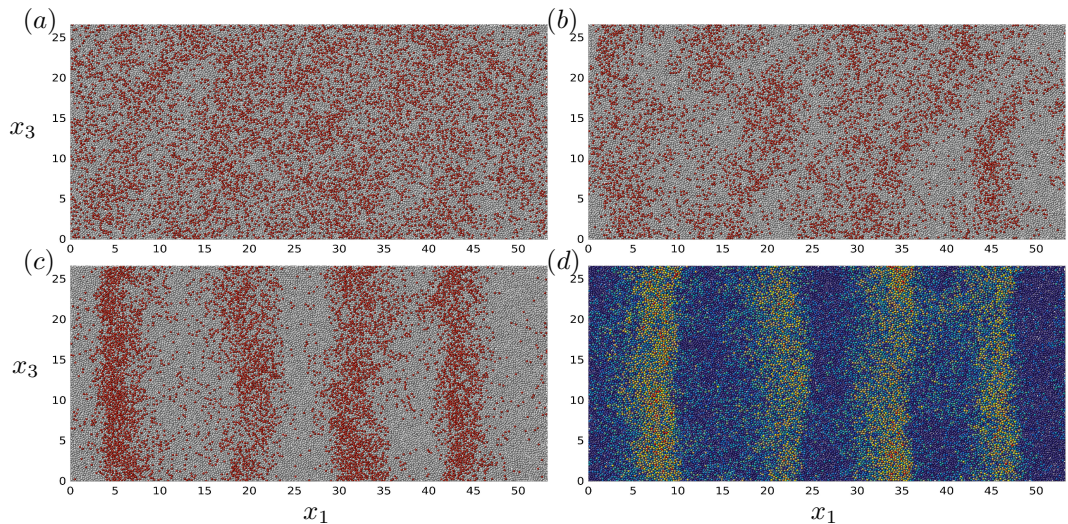
\begin{figure}[t]%[ht]
\begin{picture}(0,200)(0,0)
  \ifnine
  \put(0,94){
    \put(0,0){\includegraphics[trim=0cm 0cm 0cm 30cm, clip, width=.51\textwidth]{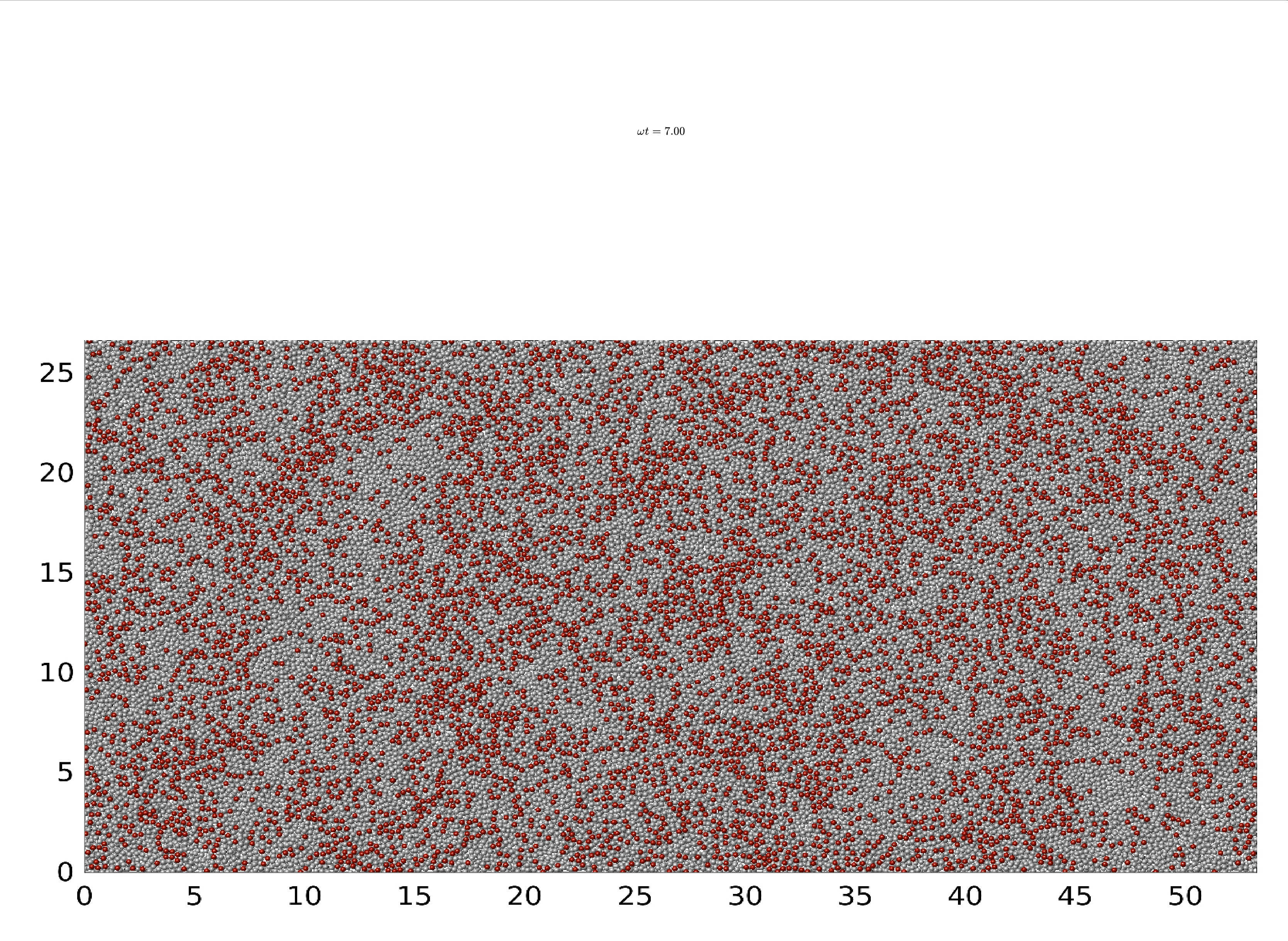}}
    \put(-10,50){$x_3$}
    \put(-5,90){$(a)$}
  }
  \put(0,3){
    \put(0,0){\includegraphics[trim=0cm 0cm 0cm 30cm, clip, width=.51\textwidth]{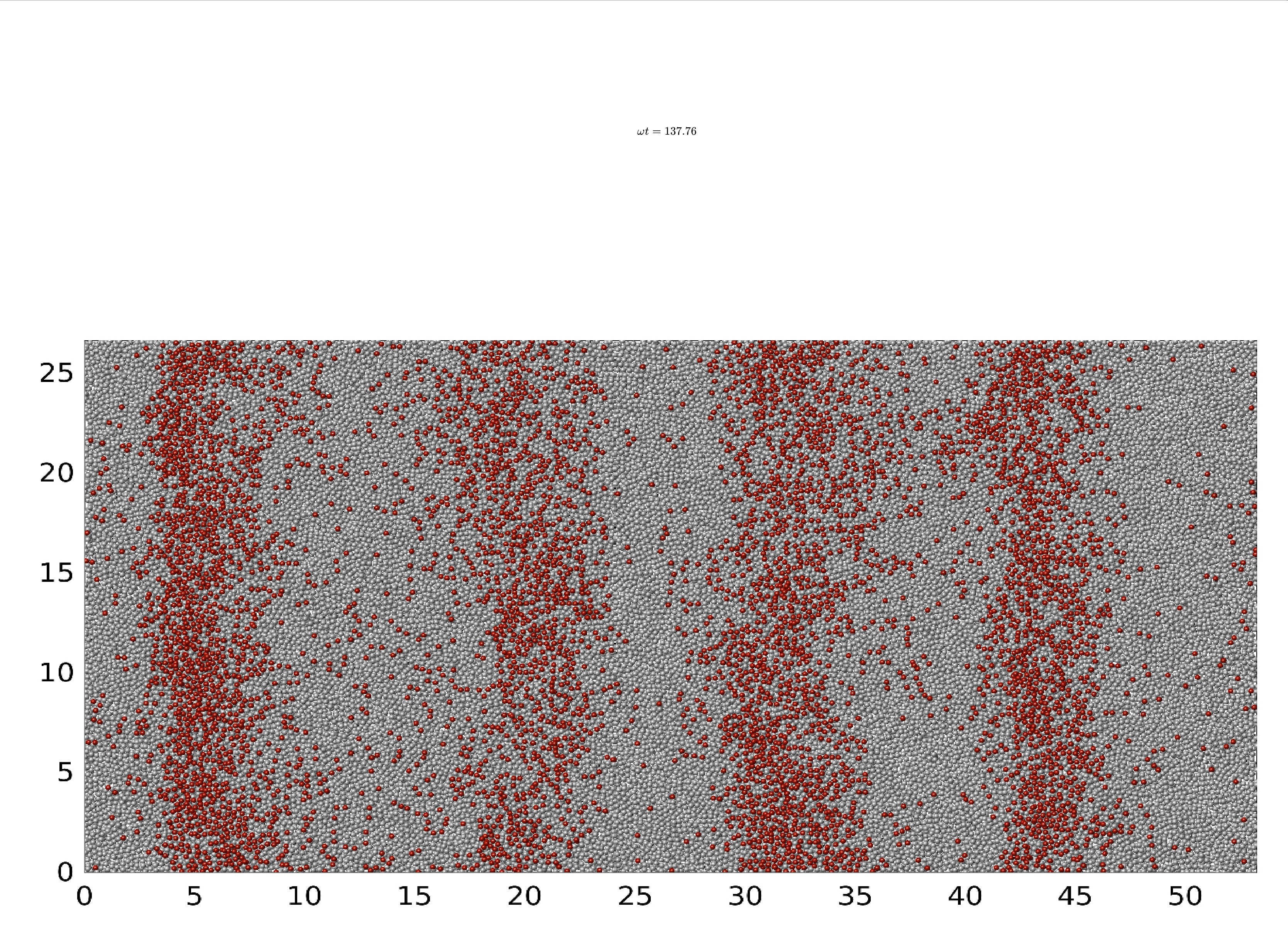}}
    \put(100,-4){$x_1$}
    \put(-9,51){$x_3$}
    \put(-5,90){$(c)$}
  }
  \put(194,94){
    \put(0,0){\includegraphics[trim=0cm 0cm 0cm 30cm, clip, width=.51\textwidth]{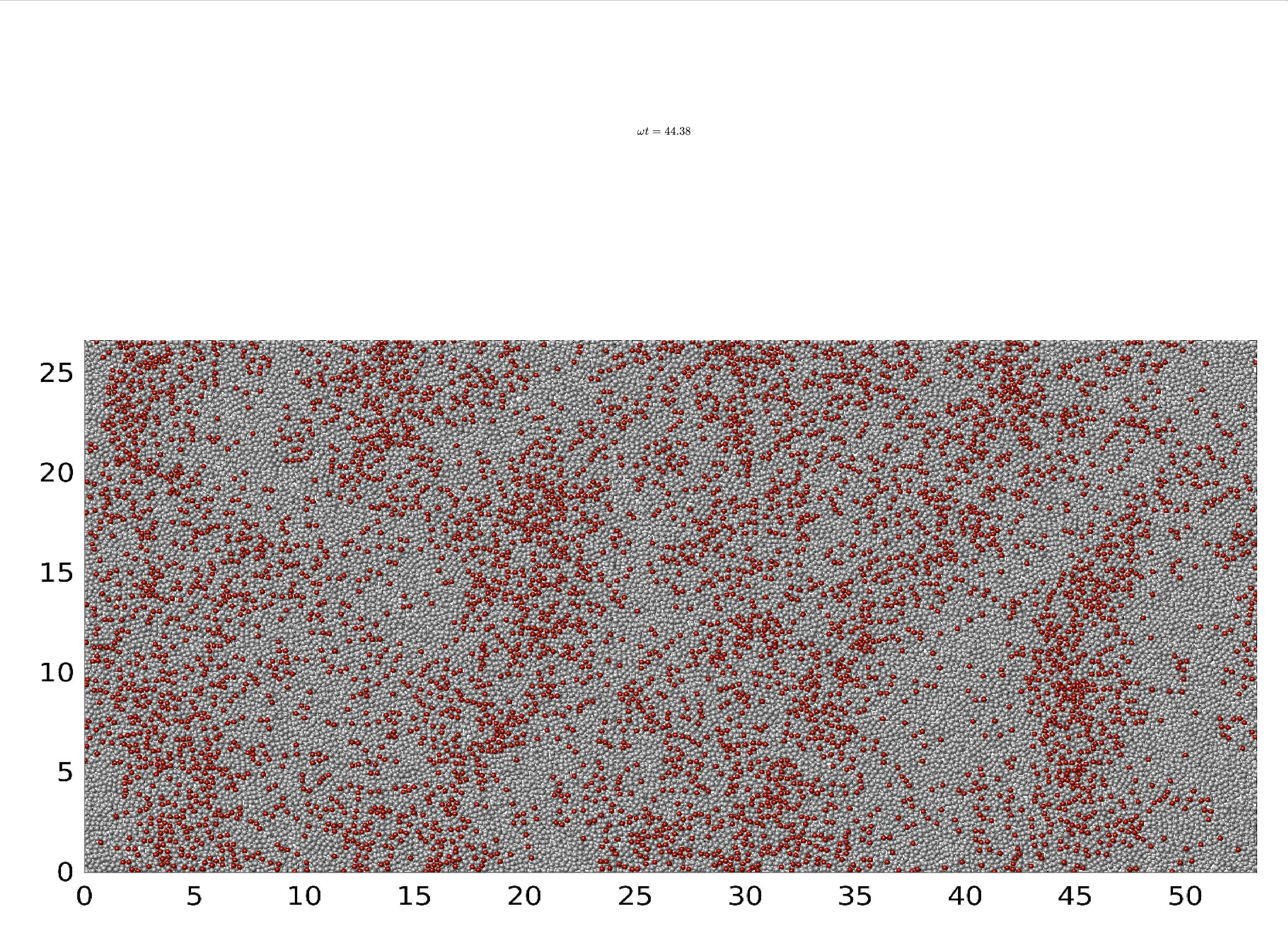}}
    \put(-3,90){$(b)$}
  }
  \put(194,3){
    \put(0,0){\includegraphics[trim=0cm 0cm 0cm 30cm, clip, width=.51\textwidth]{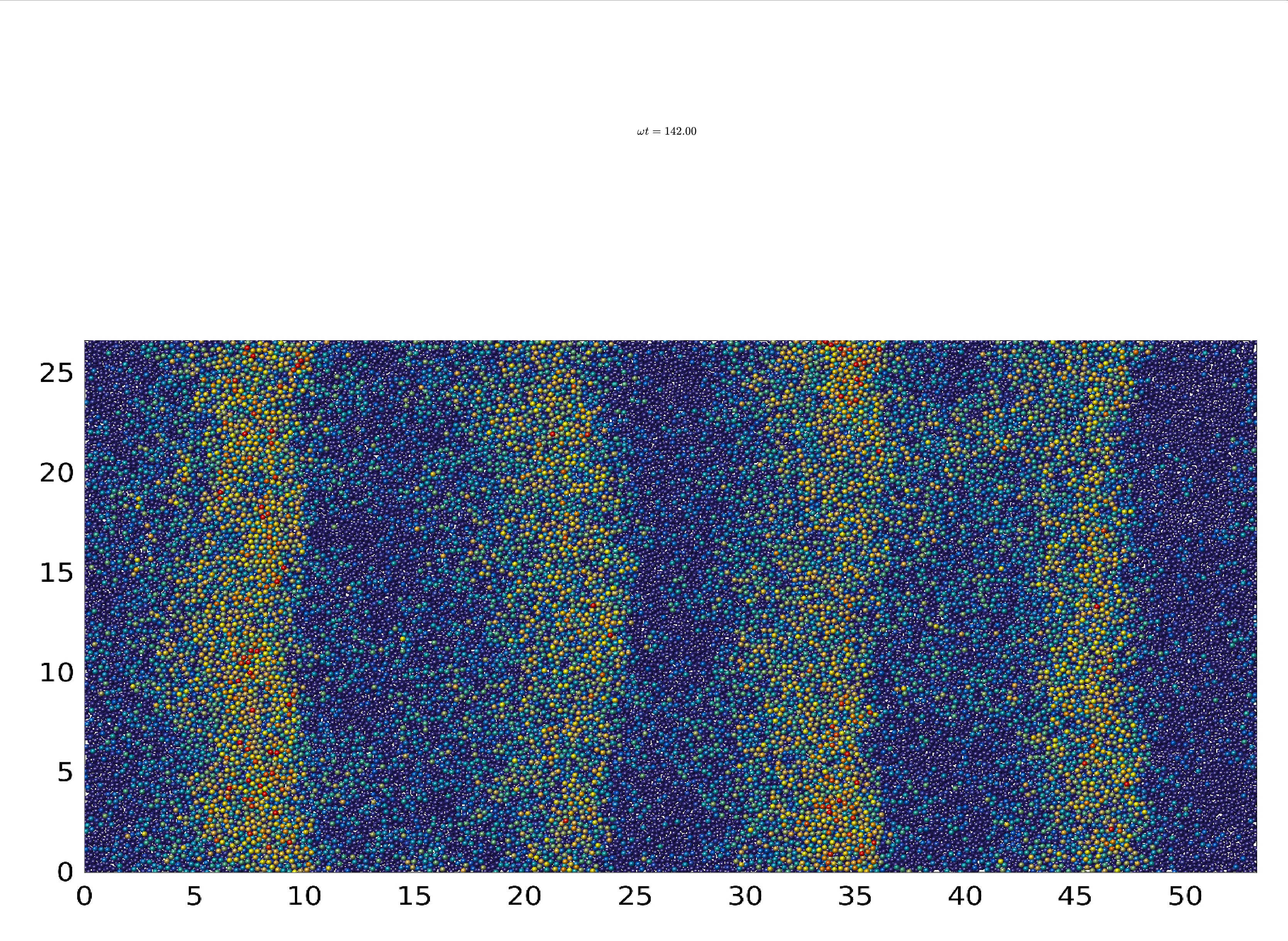}}
    \put(96,-4){$x_1$}
    \put(-3,90){$(d)$}
  }
  \fi
\end{picture}
\caption{
Top view of the bed at 
$(a)$~$t=7.0$, $(b)$~$t=44.4$, $(c)$~$t=137.8$, $(d)$~$t=142.0$ for run~$1$. Crest particles are highlighted in red in panels $(a-c)$. %
%\mmr{}{%
In panel $(d)$ particles are highlighted by colours according to their distance from the bottom, increasing from blue to red. %
%}%
%
The complete time-sequence can be seen in the movie available in the \emph{supplementary material}. %
}
\label{fig9}
\end{figure}

As mentioned above, the bed was initially leveled %(numerically) 
in order to start 
the simulations with a plane-bottom configuration. %
%
%Therefore, 
Let the spheres farthest from the %
\mmr{%
wall, %
}{%
bottom, %
}% 
%most exposed to the flow, 
i.e. whose center is located 
in the range of one diameter below the farthest one, 
be hereafter %termed 
referred to as 
\textit{crest particles} (cf. black spheres in figure~\ref{fig21}). %
Initially, crest particles 
are distributed approximately 
randomly on the bed, as shown by the red spheres in figure \ref{fig9}a. %
Then, after a few oscillations, crest particles tend to group in short chains or small bunches 
during half-periods which can eventually be destroyed in the subsequent half-period or 
merge %
\mmt{with} %
each other (see figure \ref{fig9}b). %
%%
%%    Figure 9
Finally, clear two-dimensional patterns form which then accrete and form the rolling-grain ripples (figure \ref{fig9}c,d). %
A movie of the formation of ripples for the run~$1$ can be found online as \emph{supplementary material}. %

For the present values of the parameters, only the surficial spheres %at the bed surface 
%most exposed to the flow 
exhibited %rolling motion and 
significant displacements through rolling motion while no particles 
were observed saltating or %
\mmr{entraining}{being entrained} %
into suspension. %
%
%Therefore, only bedload transport occurred. %sediment transport occurs solely in the bedload. 
%
The evolution of the bed surface and the motion of the surficial particles are described in the following. %
%
%Then the sediment flow rate is computed and related to instantaneous quantities characterising the mean flow. %
%

%%
\subsection{Evolution of the bed surface}\label{ss1}

%In two-phase Eulerian models, the 
The 
bed surface, %is commonly 
namely the (fictitious) solid-fluid interface, 
can be 
defined 
on the basis of the sediment volume concentration, hereafter referred to 
as \textit{particle volume fraction} and denoted by \svf, which %tends to vanish 
is zero 
far away from the bottom and abruptly increases %when approaching 
\mmr{on}{at} %
the bed. %\mut{CITATIONS}%
Hence, the bed surface is identified by 
points where \svf~reaches a threshold 
value. %(e.g. $\svfm=0.6$). %
Similarly, in laboratory experiments, the bed-flow interface is often detected by means of 
an image analysis procedure, thresholding the sideview frames of the bed 
\citep[e.g.][]{aussillous2013}. %
In fact, the b/w intensity of pixels is highly correlated with \svf. %
This approach was successfully reproduced numerically by \citet{kidanemariam2014} 
who considered the threshold value $\svf=0.1$. %
\citet{kidanemariam2014} defined a sample volume of size  
$\Delta x_1^*\times\Delta x_2^*\times \Lz^*$ over which the 
particle volume fraction was evaluated. 
Therefore, %$\svf\equiv\svf(x_1,t)$ 
the dependency of $\svf$ on $x_3^*$ was neglected 
%and the bed surface %coincided 
%with the %two-dimensional 
and the \emph{bed profile}, $\bprP^*(x_1^*,t^*)$, was obtained. %
%

%Such a methodology requires the definition of a sample volume over which the 
%particle volume fraction is computed that, in the case of \citet{kidanemariam2014}, 
%degenerated into spanwise %``spaghetti'' 
%elements 
%of volume 
%$\Delta x_1\times\Delta x_2\times L_{x_3}$. %
%
%This approach was also used to study the evolution of two-dimensional 
%bedforms in the present simulations. %

\begin{figure}
\begin{picture}(0,168)(0,0)
  \ifzero
  \put(44,0){
    \putpic{0,0}{.7}{figure3}
  }
  \fi
\end{picture}
\caption{
Comparison between the bed profiles $\bprE$ (red line) and $\bprP$ (black line) at an instant of run no.~$1$ when rolling-grain ripples are present.
}
\label{fig0}
\end{figure}
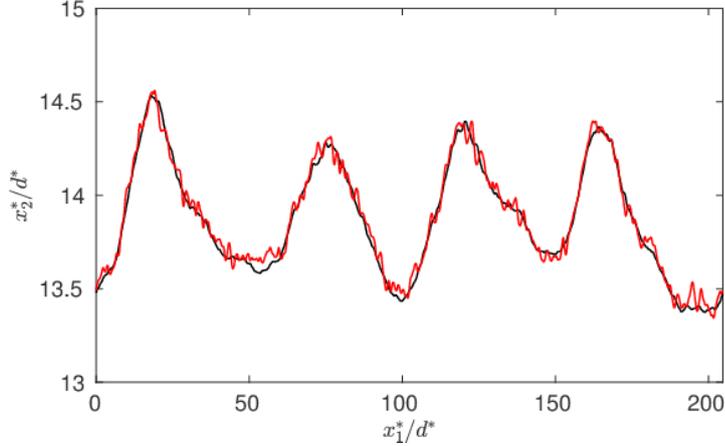

%However, also a
Another approach is also presently considered which was first adopted 
by \citet{mazzuoli2017b} to detect the bed/flow interface. %
Since spheres are presently not set into suspension, %by vortex structures, 
%their contact between them is enduring 
they remain in enduring contact 
throughout their motion and 
%the enduring contact between them persists during their motion, 
the bed surface can be thus unambiguously identified by the centers of the spheres 
on top of others, which are hereafter referred to as \emph{top-layer particles} 
(cf. dark-grey spheres in figure~\ref{fig21}). %
The $i$-th sphere ($i=1,\,\ldots,\Npt$) belongs to the top layer 
if no other sphere centers above the $i$-th one lie inside the solid angle of magnitude $\Omega=(2-\sqrt{3})\pi$~sr with respect to the % 
\mmr{%
wall %
}{%
bottom %
}%
normal. %
Then, if this condition is fulfilled, the Boolean function ${\cal E}_i$ 
associated with the $i$-th particle is equal to $1$, otherwise it is equal to $0$.
Therefore, the bed surface is defined as the function interpolating 
the centers of the spheres characterised by ${\cal E}=1$, and is 
%hereafter 
denoted by $\bsur^*(x_1^*,x_3^*,t^*)$. %
Such definition circumvents the matter of defining a threshold and allows us 
also to study three-dimensional patterns. %
Actually, the patterns observed for the present values of the parameters 
do not show an appreciable dependency on the spanwise coordinate 
%are two-dimensional (i.e. independent of the spanwise coordinate) 
and the bed profile $\bprE^*$, defined as equal to $\left\langle\bsur^*\right\rangle_{x_3}$, 
%, hereafter indicated with $\bprE$, 
is found practically to collapse on $\bprP^*$ 
once it is shifted vertically upward by a constant value $\sim 0.8 d^*$ (cf. figure~\ref{fig0}). %
Since the position of the bed profile is analogously detected by the two procedures, 
henceforth the profile $\bprP^*$ is considered. %

%%
%%    Figure 0

\begin{figure}[ht]
\begin{picture}(0,262)(0,0)
  \iffifteen
  \put(5,-3){
  \put(-14,-2){
    \putpic{0,0}{.53}{figure4a}
  }
  \put(-14,243){\small $(a)$}
  \put(158,253){\small $\bprP'^{*}/\ds^*$}
  \put(190,-2){
    \putpic{0,0}{.512}{figure4b}
  }
  \put(188,243){\small $(b)$}
  \put(352,253){\small $\bprP'^{*}/\ds^*$}
  }
  \fi
\end{picture}
\caption{
 Spatio-temporal development of the fluctuations of the bed profile about the average bed elevation, $\bprP'^{*}/\ds^*$, for run~$1$ $(a)$ and run~$2$ $(b)$. 
}
\label{fig15}
\end{figure}
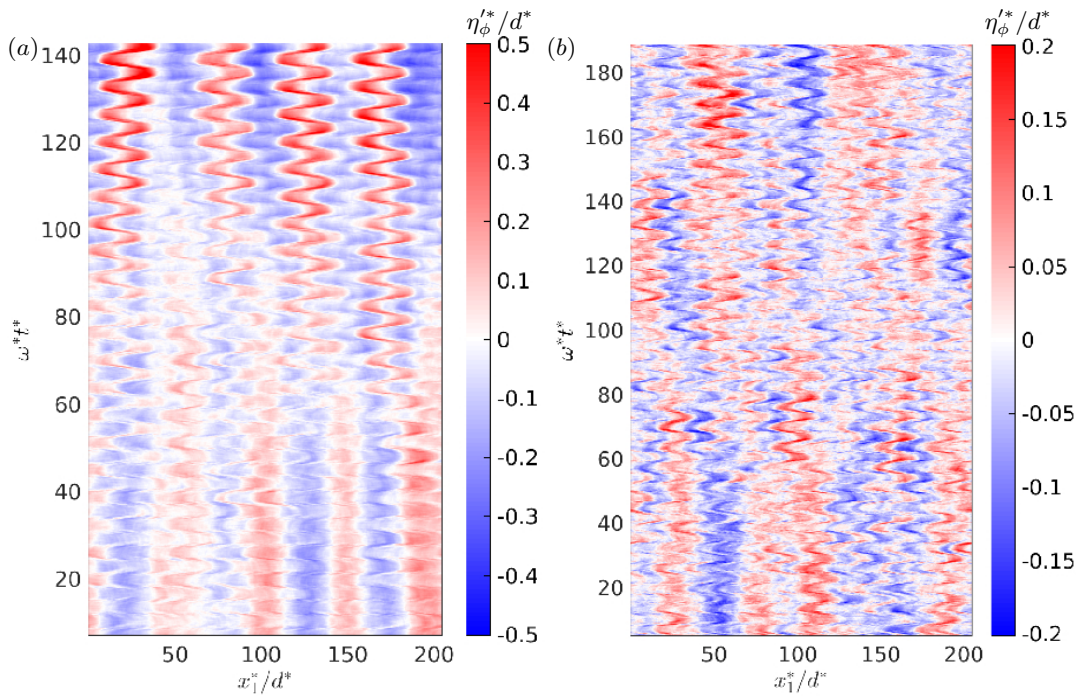

In the simulations, the average bed elevation $\sav{\bprP}_{x_1}$ initially decreases as an effect of the %settlement of particles and 
settlement and compaction of the granular bed (not shown here). %
Then, after a few oscillation periods $\sav{\bprP}_{x_1}$ asymptotically reaches a constant value %, except for 
%\mmr{}{%
approximately equal to $13.83~\ds^*$ and $12.12~\ds^*$ for both runs~$1$ and $2$, respectively, 
superimposed only by 
small fluctuations %associated with the 
of order $\mathcal{O}(10^{-2})~\ds^*$, 
which are related to the different 
phases of the wave cycle. %
%
%\mmr{}{%
Correspondingly, the solid volume fraction, $\svf$, in the region between the bottom and the surface layers of particles does not show significant temporal fluctuations and attains the average values $0.49$ for both runs. %runs~$1$ and $0.489$ for runs~$2$. %
%}%
%
Instead, figure~\ref{fig15} shows the space-time development of the fluctuations of the bed profile about the average bed elevation, i.e. $\bprP'=\bprP-\sav{\bprP}_{x_1}$, for runs~$1$ and $2$. 
%%
%%    Figure 15
%%
%%    Figure 12
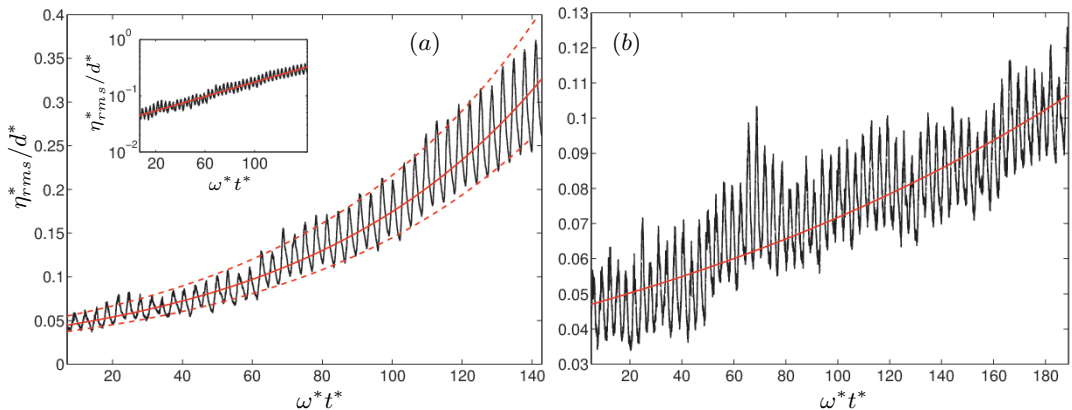
\begin{figure}[ht]
\begin{picture}(0,150)(0,0)
  \iftwelve
  \put(-2,5){
    \putpic{0,0}{.5}{figure5a}
    \put(98,-10){\small $\omega^* t^*$}
    \put(-10,65){\small \rotatebox{90}{$\eta^*_{rms}/d^*$}}
    \put(140,125){\small $(a)$}
    \put(65,72){\scriptsize $\omega^* t^*$}
    \put(19,94){\scriptsize \rotatebox{90}{$\eta^*_{rms}/d^*$}}
  }
  \put(194,5){
    \putpic{0,0}{.5}{figure5b}
    \put(98,-10){\small $\omega^* t^*$}
    \put(20,125){\small $(b)$}
  }
  \fi
\end{picture}
\caption{
%$\bprP=a\exp{(b\,t)}$, run~$1$: $a=0.0403$ $b=1.464\cdot10^{-2}$; $a_{min}=0.0338$ $b_{min}=1.455\cdot10^{-2}$; $a_{max}=0.0498$ $b_{max}=1.468\cdot10^{-2}$ (growth rate of fluctuations is approximately equal to $(a_{max}-a_{min})/a=0.3967$); run~$2$: $a=0.046$ $b=4.45\cdot10^{-3}$.
The diagrams in panel~$(a)$ and $(b)$ show the root mean square of the spatial fluctuations of the bed profile plotted versus time for simulations~$1$ and $2$, respectively. 
The thick solid (red) lines are the regression curves $a\exp{(b\,t)}$, where $a=0.040$, $b=1.46\cdot10^{-2}$ for run~$1$ and $a=0.046$, $b=4.45\cdot10^{-3}$ for run~$2$. %
\mmt{Dashed} (red) lines in panel~$(a)$ are obtained for $a=a_{min}=0.034$ and $a=a_{max}=0.050$. %
\mmt{%
  The inset figure in panel~$(a)$ highlights the exponential trend of $\bprPrms$ using 
  semi-logarithmic axis scale. %
}%
}
\label{fig12}
\end{figure}
While bedforms emerge in the second half of %
\mmr{the simulation}{run}~$1$, in %
\mmr{the simulation}{run}~$2$ %
the presence of persistent patterns is difficult to %
\mmr{be detected}{detect} %
by visual inspection of figure~\ref{fig15}b. %
%
%Indeed, the \textit{root mean square} (rms) of $\bprP$ shows that the amplitude of the fluctuations monotonically increases with time also in the latter case (see figure \ref{fig12}b), whilst 
%the value attained at the end of the simulation~$2$ 
\mmc{Indeed,} The \textit{root mean square} (rms) of $\bprP'$, $\bprPrms$, %
\mmr{amplifies}{increases} %
with time in both runs (cf. figure~\ref{fig12}), whilst the amplitude of the fluctuations attained at the end of %
\mmr{the simulation}{run}~$2$ %
%their amplitude 
barely reach $0.1~d^*$ (three times smaller than that of run~$1$). %
% at the end of the simulation $2$.
%
In run~$1$, the 
%rms of \bprP fluctuations, which can be considered perturbations of the bed surface, 
linear regression of $\ln\bprPrms$ (red solid line in figure~\ref{fig12}a) shows that $\bprPrms$ 
grows exponentially. 
Moreover, the regression of relative maxima and minima of $\bprPrms$ computed for each half-period (indicated with \mmt{dashed} lines in figure~\ref{fig12}a) preserves the exponent of the mean trend. Thus, the amplification of half-period fluctuations of the bed surface, %within each half oscillation period, 
normalised by the particle diameter, can be approximated by the expression:
\begin{equation}
\mathcal{A}_\eta=(a_{max}-a_{min})e^{b\,t}
\label{eq:rms}
\end{equation}
where $b=1.46\cdot10^{-2}$ and the factors $a_{max}=4.98\cdot10^{-2}$ and $a_{min}=3.38\cdot10^{-2}$ refer to the average upper and lower bounds of \bprPrms. %
In other words, the rate of coarsening of ripples %, at least for the first $20$ oscillation periods, 
is directly proportional to the amplitude of ripples. %which can be regarded as small perturbations of the bed surface. 
On the other hand, in run~$2$, the %growth of the fluctuations of the bed profile 
time development of $\bprPrms$ is not monotonic and the mean growth is 
slower than that observed for simulation~$1$ (cf. figure~\ref{fig12}b). %
It is likely that, in a much larger number of oscillation periods, the formation of ripples 
could be observed more clearly also in case~$2$, but this would require formidable 
computational and wall-clock time which are at the moment out of reach. %
%
%%
%%    Figure 14
\begin{figure}[ht]
\begin{picture}(0,150)(0,0)
  \iffourteen
  \put(-2,5){
    \putpic{0,0}{.5}{figure6a}
    \put(98,-10){\small $\omega^* t^*$}
    \put(-10,65){\small \rotatebox{90}{$\vert\hat{\eta}^*\vert/d^*$}}
    \put(22,125){\small $(a)$}
  }
  \put(194,5){
    \putpic{0,0}{.5}{figure6b}
    \put(98,-10){\small $\omega^* t^*$}
    \put(22,125){\small $(b)$}
  }
  \fi
\end{picture}
\caption{
Panels $(a)$ and $(b)$ show the absolute value of four Fourier modes of the bed profile plotted versus time for simulations~$1$ and $2$, respectively. Red, blue, black and magenta lines correspond to the $2_{nd}$, $3_{rd}$, $4_{th}$ and $5_{th}$ modes of the bed profile, respectively. 
}
\label{fig14}
\end{figure}
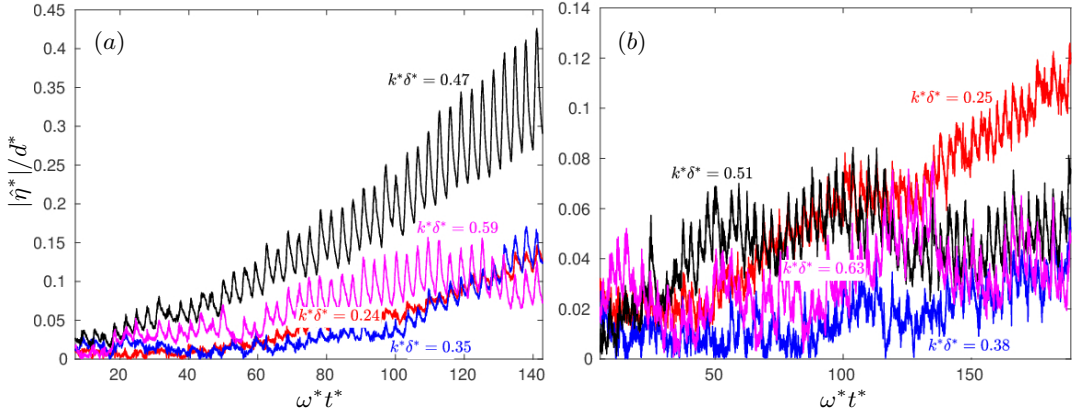
%%

%There is more than one possibility to detect the wavelength of ripples as a function of time. %
%It is also possible to 
%An approach t
To detect the wavelength of ripples as a function of time, %consists in expanding 
the bed profile is expanded in Fourier series and 
the absolute value and the growth rate of each term of the series is investigated. %
%
%The time development of the absolute value of the Fourier coefficients related to the wave 
%numbers that give a significant contribution to the series is shown in figure~\ref{fig14}. %
%
\mmc{Indeed,}\mmt{It is evident that} the wave numbers $k^*\del^*=0.47$ and $k^*\del^*=0.25$ dominate the spectra of $\bprP'$ at the end of runs~$1$ and $2$, respectively (cf. figure~\ref{fig14}). %
However, since 
%the evolution of each wave number is not independent of the others and the dependence is in general nonlinear it is hard to estimate the growth function associated with each mode.
modes are still evolving at the end of each run, an equilibrium condition is not reached and 
the simulation time is not sufficient to describe the complete evolution of individual modes. %
%
%A suitable approach consists in defining, as a function of time, 
Alternatively, the \textit{dominant wavelength} can be defined as two times the space lag, 
$\wln^*$, at which the absolute value of the two-point correlation function of 
$\bprP'^*$ attains the first maximum value \citep{kidanemariam2017}. %
%
%%
%%    Figure 13
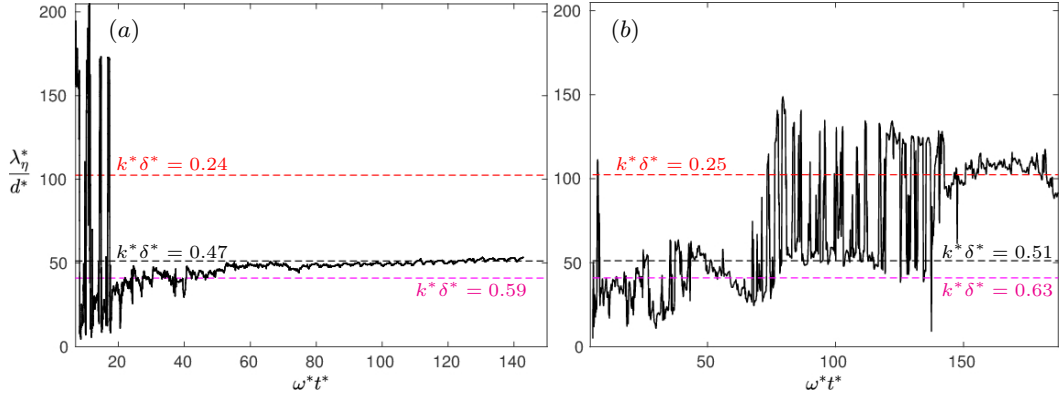
\begin{figure}[ht]
\begin{picture}(0,150)(0,0)
  \ifthirteen
  \put(186,2){
    \putpic{0,-5}{.52}{figure7a}
  }
  \put(-7,2){
    \putpic{0,-5}{.525}{figure7b}
  }
  \put(30,130){\small $(a)$}
  \put(33,79.5){\color{red!}\scriptsize $k^*\del^*=0.24$}
  \put(33,46.5){\color{black!}\scriptsize $k^*\del^*=0.47$}
  \put(145,32){\color{magenta!}\scriptsize $k^*\del^*=0.59$}
  \put(218,130){\small $(b)$}
  \put(220,79.5){\color{red!}\scriptsize $k^*\del^*=0.25$}
  \put(342,46.5){\color{black!}\scriptsize $k^*\del^*=0.51$}
  \put(342,32){\color{magenta!}\scriptsize $k^*\del^*=0.63$}
  \fi
\end{picture}
\caption{
Dominant wavelength of the bed profile computed as a function of time for %
\mmr{the simulations}{runs}~$1$ %
(panel $(a)$) and $2$ (panel $(b)$).
}
\label{fig13}
\end{figure}
The result of this procedure is shown in figure~\ref{fig13}. %
\mmc{Indeed,} As predicted by the Fourier analysis, the dominant wavelength for the second half 
of simulation~$1$ corresponds to the wave number $k^*\del^*=0.47$ ($\wln^*=13\delta^*$) 
while for the last $\sim 6$ oscillation periods of simulation~$2$ patterns are characterised 
by the wave number $k^*\del^*=0.25$ ($\wln^*=25\delta^*$). %are dominant. %
The values of $\wln^*$ can be compared with the results of the experiments carried out by 
\citet{blondeaux1988} for similar values of the parameters and with those obtained 
by linear stability analysis by \citet{blondeaux1990}. %
It is found that %$\wln^*$ 
%mu
%has the same magnitude of the wavelength of the first emerging ripples observed in the laboratory. %
the values of $\wln^*$ in the current simulations %
\mmr{have values}{are} %
comparable to the wavelengths of 
the first emerging ripples observed in the laboratory. %
At this stage it is worthwhile to remark %
\mmt{on} %
the importance of this result, since a natural 
very complex phenomenon has been reproduced by a very simplified, though numerically 
challenging, system, which indicates that the basic process leading to the formation 
of ripples is somewhat robust. %
In particular, for the experiment reproduced by run~$1$, \citet{blondeaux1990} observed 
$\wln^*=25\delta^*$ ($96~d^*$) while for the case simulated by run~$2$ the value of $\wln^*$ 
was approximately equal to $26\delta^*$ ($108~d^*$). %
\mmr{Analogous}{Similar} %
results are predicted by means of the linear stability analysis following the approach of \citet{blondeaux1990} ($\wln^* = 23\delta^*$ for run~$1$ and $\wln^*=22\delta^*$ for run~$2$). %
\citet{rousseaux2004} carried out experiments also exploring the region of the parameter space where %
\mmr{simulations}{runs}~$1$ %
and $2$ lie and observed the first measured wavelengths $\wln^*\sim 20~\delta^*$ and $\wln^*\sim 25~\delta^*$, respectively. %
Therefore, the wavelength of ripples simulated in %
\mmr{the case}{run}~$1$ is smaller than the wavelength observed experimentally.
Such a discrepancy can be due to several reasons mostly associated with the modelling of 
particle-particle interactions. %
%One of them is presumably associated with the fact that 
%\mmr{}{%
The significance of the role of sediment friction in the formation of patterns 
was emphasised by \citet{moon2004}. %
Indeed, sand grains can have irregular shape and, consequently, more than one point of contact 
during a binary collision, which allow them to transfer linear and angular momentum more 
efficiently than spheres. %
%} %
%For instance, since sediment particles are approximated by spheres, the contact is different from 
%that among sand grains which have irregular shape. %
%
Moreover, the sensitivity of sediment dynamics to the contact is
enhanced if particles roll over each other (enduring contact) rather than colliding. %
The particles of run~$1$ behave like finer sand grains, 
since, %
\mmr{on the basis of}{based on} %
the experimental results of \citet{rousseaux2004}, 
the first measured wavelength tends to increase %if sediments of larger size are considered
monotonically with increasing %
\mmr{the}{} 
size of sediments 
%\citep[e.g.][]{rousseaux2004} 
and because spherical particles are statistically set into motion more easily and %
\mmr{perform}{undergo} larger excursions than sand grains of irregular shape. %and equivalent size, 
%in the case of simulation~$1$.
%
%It is likely that ripples would merge two-by-two if the simulation~$1$ was continued, thereby $\wln^*$ attaining the same value as that observed experimentally. 
%
%In fact, as shown in 
On the other hand, figure~\ref{fig14}a shows that the modes associated with the wave numbers $k^*\del^*=0.35$ and $k^*\del^*=0.24$ grow 
\mmt{at} %
approximately %
\mmr{with}{} % 
the same rate as the dominant mode 
in the last periods of %
\mmr{the simulation}{run}~$1$ %which make us presume 
and it is possible 
that the four ripples of 
figure~\ref{fig9} might merge %two-by-two 
after a certain time. %
%the simulation was continued. %, thereby $\wln^*$ attaining values approximately equal to that observed experimentally. 
%

%%
%%    Figure 16
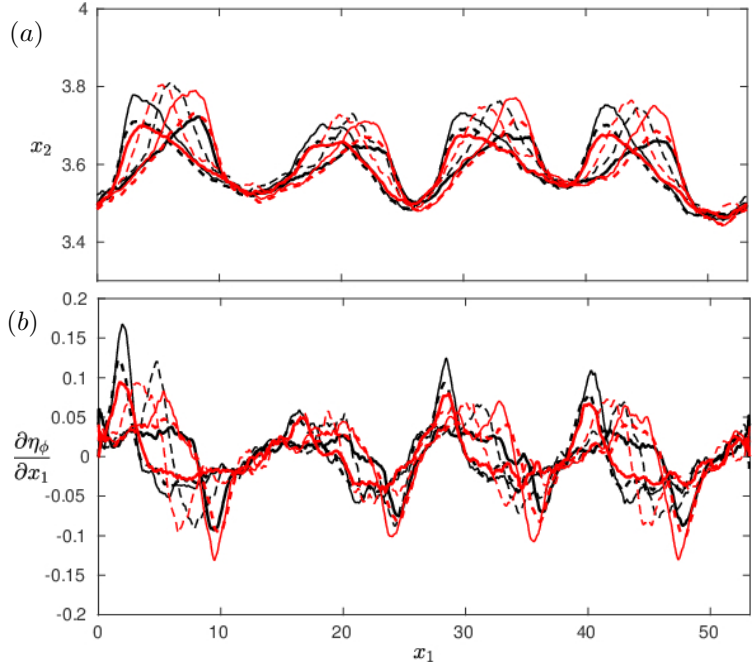
\begin{figure}[ht]
\begin{picture}(0,245)(0,0)
  \ifsixteen
  \put(41,124){
    \putpic{0,0}{.7}{figure8a}
    \put(-8,110){$(a)$}
  }
  \put(35,0){
    \putpic{0,0}{.718}{figure8b}
    \put(-3,125){$(b)$}
  }
  \fi
\end{picture}
\caption{
Bed profile, $\bprP$, (panel $(a)$) and bed slope (panel $(b)$) at different phases of a wave cycle when rolling-grain ripples are formed. Black lines refer to instants $t=136.7$ (solid, thick), $t=137.4$ (\mmt{dashed}, thin), $t=138.2$ (solid, thin) and $t=139.0$ (\mmt{dashed}, thick) while the mean flow is directed from right to left. Red lines refer to instants $t=139.8$ (solid, thick), $t=140.6$ (\mmt{dashed}, thin), $t=141.4$ (solid, thin), $t=142.2$ (\mmt{dashed}, thick) during which the mean flow is directed from left to right.
}
\label{fig16}
\end{figure}
%%
%%    Figure 20
\begin{figure}[ht]
\begin{picture}(0,195)(0,0)
  \iftwenty
  \put(90,5){
    \putpic{0,0}{.5}{figure9}
    \put(-14,86){\small\rotatebox{90}{$\SSp{\bsur}^*/d^{*\,2}$}}
    \put(95,-8){\small$k^*d^*$}
    \put(25,125){\scriptsize\color{red!}$\propto (k^*d^*)^{-3}$}
    \put(64,95){\scriptsize\color{magenta!}$\propto (k^*d^*)^{-1.2}$}
    \put(120,31){\scriptsize\color{blue!}$\propto (k^*d^*)^{-4.3}$}
    \put(66,15){\small$x_1^*/d^*$}
    \put(19,45){\small\rotatebox{90}{$x_2^*/d^*$}}
    \put(69,160){\color{blue!}\line(2,1){10}}
    \put(79,164){\scriptsize\color{blue!}\textit{piecewise ramps}}
  }
  \fi
\end{picture}
\caption{
Spectra of the bed profile computed at the phases indicated in figure~\ref{fig16} for run~$1$. The solid (blue) line indicate the spectrum of the simplified configuration sketched in the small inset of the figure with periodic (blue) straight lines. 
}
\label{fig20}
\end{figure}
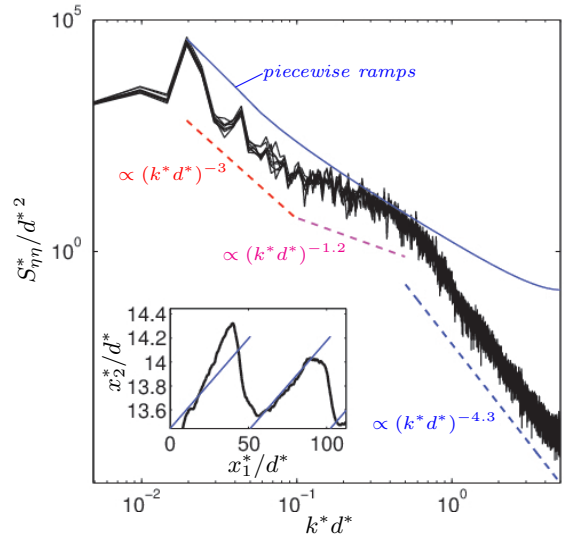
%%
%Once rolling-grain ripples form, their

%The shape of ripples changes during the oscillation period. %
%
Even though the ripples of run~$1$ do not really 
%visibly 
drift in the streamwise direction, %in the period timespan, 
their crests %undergo displacements of the order of several sphere diameters 
migrate to and fro %
\mmr{of}{by} %
several sphere diameters. %in the $x_1$-direction
Therefore, as shown in figure~\ref{fig16}, %
\mmr{the shape of ripples}{ripple shape} %
changes during the oscillation period. %
The profiles indicated in figure~\ref{fig16}a by (black and red) thick lines are attained in the 
phases when the fluid far from the bed decelerates and then vanishes %
\mmt{%
(at the flow reversal) %
}%
while surficial particles 
are at rest. %
%
%These %configurations 
%profiles 
%are %the most stable since the bed slope is relatively mild 
%characterised by relatively mild slopes 
%and persist for quite long phases (cf. figure~\ref{fig16}b). 
%
In these phases, ripples are asymmetric with slopes relatively mild and the %side with mild slope oriented in the direction of the flow. 
lee side steeper than the stoss side (cf. figure~\ref{fig16}b). %
In the subsequent phases, while the flow accelerates, the amplitude of ripples increases, along with %
\mmr{%
the bed %
}{%
their %
}%
slope, and reaches the maximum value %
\mmt{%
approximately $0.20\,\pi$
}%
earlier than the free-stream velocity does. %
Now the profile of each ripple is symmetric (thin dashed lines in figure~\ref{fig16}a), 
but then it becomes asymmetric again as the crest 
proceeds the excursion towards the other side of the ripple. %
%migrates from one side to the other. 
%
Finally, the opposite resting configuration is attained %again 
while the fluid is already decelerating. %
Hence, most of the bed-profile evolution is carried out during the acceleration phases. %
At the end of the present simulations the mild (lee) slope of ripples is approximately $0.02$ while for the steep (stoss) side the slope ranges between $0.08$ and $0.16$. 
Following the empirical approach of \citet{sleath1984}, the flow separation behind the crests should occur if 
%mu
%the average slope defined as 
the ratio between the height and the wavelength of ripples, 
namely the average steepness, 
reaches the value $0.1$. %
In the period considered in figure~\ref{fig16}, the average slope is about $0.02$ %
\mmr{, in fact}{and} %
the flow does not separate and rolling-grain ripples do not evolve into vortex ripples. %

The spectra of the bed profile, $\SSp{\bsur}^*$, are computed as functions of wave numbers at the phases of the oscillation period shown in figure~\ref{fig16} and are plotted in figure~\ref{fig20}. %
\mmr{%
The spectra of sand waves and fluvial dunes were obtained by 
\citet{hino1968,jain1974,nikora1997,coleman2011,kidanemariam2017}, 
who emphasized the fact that, for wave numbers much smaller than the smallest flow 
scale which do not affect the stability of the bed and much larger than the grain size, 
the spectrum was proportional to $k^{*\,-3}$. %
}{%
Previous research has shown that for wave numbers much smaller than the smallest flow scale which
do not affect the stability of the bed and much larger than the grain size, the spectrum was 
proportional to $k^{*-3}$ \citep{hino1968,jain1974,nikora1997,coleman2011,kidanemariam2017}. %
}%
\mmt{%
\mmc{Indeed,} \citet{hino1968} showed that, when the equilibrium configuration of the bed profile is reached, the spectrum of the bed slope depends linearly on the wave number, whence the exponent $-3$ for the spectrum of the bed profile is obtained by purely dimensional reasoning. %
From the geometrical point of view, the $-3$ power law indicates that the bed profile is self-similar, i.e. the shape of the profile is independent of the length scale \citep{nikora1997}, in the range of length scales between $\ds^*$ and $\wln^*$. %
}%
Presently, the flow is unsteady and a comparison to bedforms that reached the equilibrium configuration is not possible. %
However, in the range of wave numbers indicated by \citet{hino1968}, i.e. $0.02\lesssim k^*d^*\lesssim 0.1$ in figure~\ref{fig20}, the %
%\mmr{%
spectrum of ripple profiles is observed proportional to $k^{*\,-3}$. %
%}%
%
\mmr{Actually, t}{T}he same trend can be obtained, in this range of wave numbers, by considering the spectrum of streamwise-periodic ramps (see the inset in figure~\ref{fig20}). 
%mu
%inclined by $0.015$ (anyway the spectrum slope is independent on the slope of the ramps). %
%
Therefore, the so called ``$-3$ power-law'' is associated with the fact the the stoss side of the ripples is mostly straight. %
\mmt{%
%From the physical point of view, the exponent $-3$ indicates that bedforms have reached an equilibrium configuration, at least for a certain range of wave numbers, the spectrum of the bed slope being a linear function of such wave numbers \citep{hino1968}, %
%
%while from the geometrical point of view it indicates that the bed profile is self similar, i.e. the shape of the profile is independent of the length scale \citep{nikora1997}. %
}%
%%
%
%\mmr{%
In the laboratory or in the field, it is difficult to compute the spectra for large wave numbers because the measurements of the bed surface do not typically reach such high accuracy. %
%}%
%
For values of $ k^*d^*$ ranging between $0.1$ and $1$ (i.e. for length scales of order $\mathcal{O}(d^*)$), the slope of the spectrum is approximately equal to $-1.2$, which suggests that the fluctuations of the bed profile at these scales are nearly random (i.e. the scales in this range are uniformly present). %
Consequently, in this range the spectrum of the bed slope, which is equal to 
$2\pi k^{*\,2}\SSp{\bsur}^*$, increases with increasing values of $k^*$ and reaches a 
relative maximum at $k^*d^*\sim 0.5$. %
In other words, most of the fluctuations of the bed slope in such range of wave numbers are 
characterised by the length scale $2\,d^*$. %
\mmt{%
\citet{nikora1997} showed that such ``bulges'' of the bed profile spectra are associated with scale transitions, for instance between the meso- and micro-scales. % 
}%
Finally, for values of $k^*d^*$ larger than $1$ the spectrum decreases with slope $-4.3$, which was also observed by \citet{kidanemariam2017}. %
The latter range is not relevant for the characterization of the bedform geometry and the trend of the spectrum is possibly related to the shape of sediment particles. %
%

%%
%%    Figure 17
\begin{figure}[ht]
\begin{picture}(0,170)(0,0)
  \ifseventeen
  \put(-2,-5){
    \putpic{0,0}{.5}{figure10a}
    \put(0,160){$(a)$}
  }
  \put(200,-2){
    \putpic{0,0}{.48}{figure10b}
    \put(0,118){$(b)$}
  }
  \fi
\end{picture}
\caption{
Steady streaming visualised by means of streamlines of the spanwise and time averaged flow field. The thick solid %
\mmt{(red)} %
lines indicates the average bed profile. Panels $(a)$ and $(b)$ refer to runs~$1$ and $2$.
}
\label{fig17}
\end{figure}
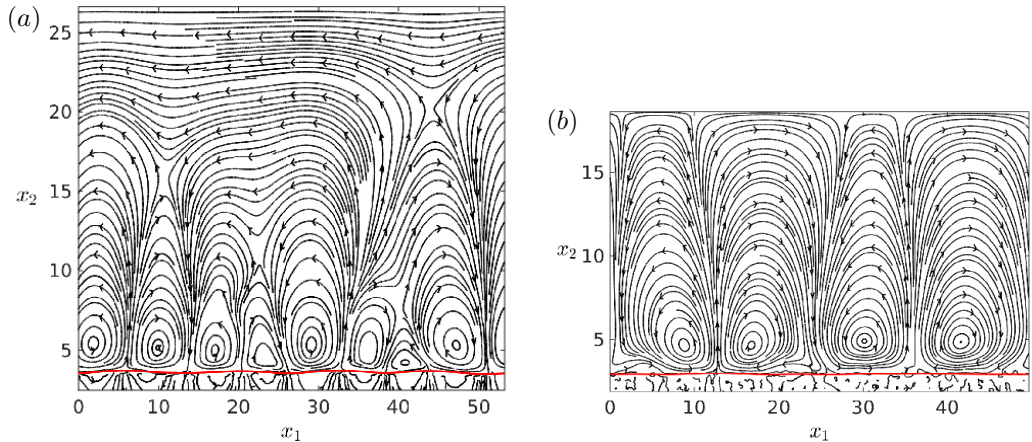
%%

%\mmr{}{%
\citet{mazzuoli2016a} %
\mmr{pointed out}{showed} %
that the interaction between a few spheres rolling over a plane % 
\mmr{%
wall %
}{%
bottom %
}%
and the oscillatory flow, promoted the growth of certain disturbances and the decay of others, independently of the presence of roughness elements. 
Indeed, the formation of ripples is strictly related to the development of steady 
streaming. 
Presently, two-dimensional recirculating cells originated over the bed surface after 
a few oscillation periods. %
%%
%%    Figure REV1
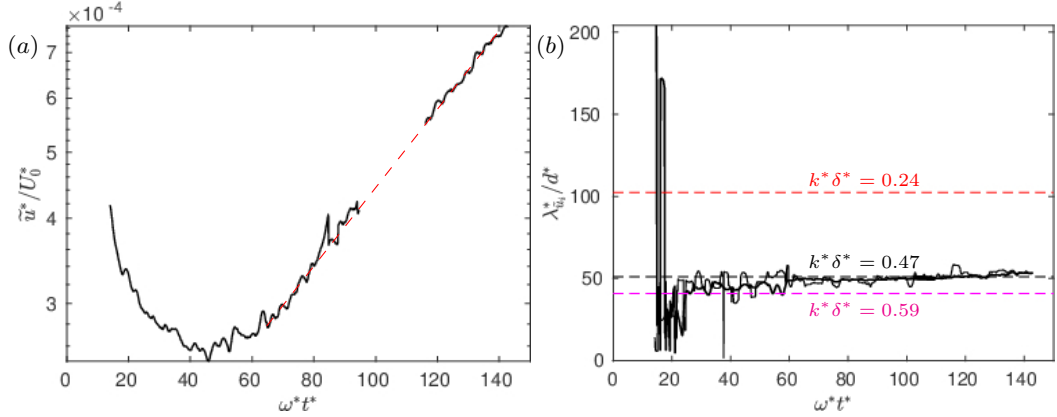
\begin{figure}[ht]
\begin{picture}(0,155)(0,0)
  \put(-2,-2){
    \putpic{0,0}{.5}{figure11a}
    \put(-4,135){\small $(a)$}
    \put(93,32){\color{red!}\rotatebox{52}{\multiput(0,0)(8,0){18}{\line(1,0){4}}}}
  }
  \put(194,-2){
    \putpic{0,0}{.5}{figure11b}
    \put(-2,135){\small $(b)$}
    \put(100,85){\color{red!}\scriptsize $k^*\del^*=0.24$}
    \put(100,54){\color{black!}\scriptsize $k^*\del^*=0.47$}
    \put(100,36){\color{magenta!}\scriptsize $k^*\del^*=0.59$}
  }
\end{picture}
\caption{
\mmt{%
Panel $(a)$ shows the magnitude of the steady streaming averaged over the $x_1x_2$-plane, $\sstr^*$, normalised by $U_0^*$ and plotted versus time with the vertical axis in logarithmic scale. %
In panel $(b)$ the dominant wavelength of the streamwise (thick line) and wall-normal (thin line) components ($i=1,2$) of the steady streaming velocity at $x_2^*=3.90\delta^*$. 
Ripples appear nearly when the growth rate attains an exponential trend with constant rate.
Each point of the curve is referred to the time-averaged value computed over the previous period for run~$1$. %
}% mmt
}
\label{fig:steadystream}
\end{figure}
%%
%}%
%
In the final part of the simulations, the steady streaming appears as in figure~\ref{fig17} 
which was obtained by averaging the flow field in the spanwise direction and over the 
last $3$ periods of run~$1$ and the last $5$ periods of run~$2$. %
\mmt{%
Let the intensity of the steady streaming, $\sstr^*(t)$, be defined as the magnitude 
of the average in the interval $[t-\TT,t]$ and in the spanwise direction of the flow field, namely 
$\sstr^* 
= 
\sav{(
\sstr^{*2}_1 
+ 
\sstr^{*2}_2
)^{1/2}}_{x_1x_2}$, 
where $\sstr^*_i\equiv\sav{u_i}^*_{T,x_3}$ is the $i$-th component of the period-and-spanwise averaged fluid velocity and $i=1,\,2$. %
Figure~\ref{fig:steadystream}a shows that, in run~$1$, the value of $\sstr^*/\U^*$ initially decreases, then starts to increase and attains an exponential growth from approximately the $10$th oscillation period on, similarly to ripples of wavelength $\wln$. %
%
%Simultaneously, ripples of wavelength $\wln$ form. %
%
%Thus, both the rms of bed surface fluctuations, $\bprPrms$, and the intensity of the recirculating cells show a similar evolution. %
%
Moreover, the evolution of dominant wavelength of the bed profile in figure~\ref{fig13}a matches closely the evolution of $\wlns$, namely the dominant spatial periodicity of $\sstr_1$ and $\sstr_2$ in the vicinity of the bed, which is shown in figure~\ref{fig:steadystream}b. %
Hence it is evident that the formation of ripples is coupled with the development of recirculating cells. %
}%mmt
The maximum velocity of the steady streaming is attained in the vicinity of the bed 
surface and is approximately equal to $1.5\cdot10^{-2}~\U^*$ and $0.6\cdot10^{-2}~\U^*$ 
for runs~$1$ and $2$, respectively. %
Close to the bed, the spatial periodicity and the flow direction of recirculating 
cells promote the accretion of the ripples characterised by wavelength equal to $\wln^*$. %
In particular, $4$ pairs of recirculating cells can be observed for run~$1$ 
and $2$ pairs for run~$2$. %
However, figure~\ref{fig17}a, which refers to %
\mmr{the simulation}{run}~$1$, shows that 
recirculating cells with different periodicity superimpose above the bed and at 
$x_2 = 13$ only $2$ pairs of recirculating cells can be detected. %
This is compatible with the %prediction formulated above that $\wln$ might increase in many oscillation periods. %
evolution of the bed profile described above, in particular with the 
growth of the mode characterised by $k^*\del^*=0.24$, as shown in figure~\ref{fig14}a. %
%
%Such a rich picture is absent in the figure~\ref{fig17}b referred to the other simulation. 
In run~$2$, contrarily, recirculating cells do not merge far from the bed (cf. figure~\ref{fig17}b). %

\subsection{Dynamics of surficial particles}\label{ss2}

\mmr{%
So far the process generating the ripples has been described which appears 
essentially driven by the steady secondary flow arising in the boundary layer. %
We now wonder what is the role that moving particles play in the coupled problem of 
the bed surface evolution. %
}{%
The process generating the ripples has been described as primarily being driven by the 
steady secondary flow arising in the boundary layer. %
In this section, we will evaluate the role that moving particles play in the coupled problem 
of the bed surface evolution. %
}%
\citet{blondeaux1990} found that the first observable (often called \textit{critical}) 
wave number of ripples plotted versus $\Rdel$ 
(for fixed values of the other parameters) exhibited discontinuities whenever %
\mmr{particles, during their motion,}{particles in motion} %
interacted with a different number of recirculating cells. %
Indeed, the selection of the critical wave number 
is closely related to the ratio between the \textit{sediment semi-excursion}, $\pse^*$, 
namely the amplitude of particle oscillations in the streamwise direction, 
and the wavelength of ripples, $\wln^*$. 
%}%
However, $\pse^*$ is difficult to measure in the laboratory and \citet{blondeaux1990} 
replaced it in his study with the fluid semi-excursion, $\fse^*$, 
since the two quantities are well correlated. 
\citet{mazzuoli2016a}, who investigated by DNS the dynamics of a small %
\mmr{amount}{number} %
of spherical particles in an oscillatory boundary layer, computed the particle semi-excursion 
and found that it tended to increase almost linearly during the first oscillation periods, 
independently of the presence of bottom roughness,  
%in the case a single line of movable spheres was considered 
as long as particle-particle interactions were relatively unimportant 
\citep[tests $2$, $3$ and $4$ of][]{mazzuoli2016a}. %
The evolution of $\pse^*$ was more complex when many particles were considered.
Presently, the motion of the \textit{top-layer particles}, i.e. the spheres at the bed surface, is investigated. 
\mmr{%
Top-layer particles consist of $\mathcal{O}(2\cdot10^4)$ spheres and, 
for run~$1$ (run~$2$), 
approximately $12\%$ ($15\%$) of these %, hereafter referred to as \textit{crest particles}, 
are crest particles, i.e. lay within a distance $d^*$ 
%above the horizontal plane distant $d^*$ 
from the sphere on top of the bed. %, relatively to run~$1$ (run~$2$). %
}%
{%
Top-layer particles consist of $\mathcal{O}(2\cdot10^4)$ spheres. %
Approximately $12\%$ of these particles in run~$1$, and $15\%$ in run~$2$, are crest particles, i.e. lay within a distance $d^*$ from the sphere on top of the bed. %
}%
%
%Indeed, the selection of the first observable wave number of ripples is closely related to 
%the ratio between the sediment semi-excursion, $\pse^*$, and the wavelength of ripples, $\wln^*$ %\citep{blondeaux1990}.
%
%In order to characterise the motion of the crest particles, their 
%The trajectory of t
The top-layer particles %was computed 
are tracked 
%in order to characterise their motion 
during each 
half-period starting from the phase, $\chi$, when particle %
\mmr{%
velocity vanishes and then reverses. %
}{%
motion ceases and then restarts in the opposite direction. %
}%

\begin{figure}[ht]
\begin{picture}(0,150)(0,0)
  \ifeighteen
  \put(43,2){
    \includegraphics[width=.7\linewidth,trim=.5cm .5cm 0cm 0cm,clip]{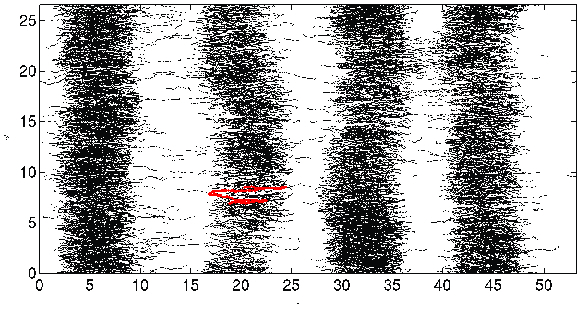}
  }
  \put(178,-4){$x_1$}
  \put(35,75){$x_2$}
  \fi
\end{picture}
\caption{\noindent%
Trajectories of crest particles %lying on the crest of ripples perform 
during the time interval $126.3<t<138.9$ ($2$ oscillation periods) of simulation~$1$. The trajectory of one particle is highlighted by a thick red line.
}
\label{fig18}
\end{figure}

In particular, the particles that at time 
%$\tin{k}=(k+\phs)~\frac{\TT}{2}$, $k=0,1,2,\ldots$ were top-layer particles, 
$\tin{j}=\pi(j+\phs)$, $j=0,1,2,\ldots$ are top-layer particles, 
are tracked for a %the following 
half-period, %
%\mmr{
$\phs$ %, is found 
being 
equal to $0.2$ for both run~$1$ and run~$2$. %
%} %
%
%the spheres lying on the crests of ripples 
Crest particles 
are more exposed to the flow, 
which gives them higher mobility than those 
lying in the troughs between the ripples. 
%than those 
%lying in the troughs between the ripples, %which give them 
%and are therefore characterised by higher mobility. 
%which allow them possibly to travel through different recirculating cells during an oscillation period. %
The trajectories of crest particles obtained for two periods of simulation~$1$ (when four ripples are present) are marked in figure~\ref{fig18}. 
%%
%%    Figure 18
Figure~\ref{fig18} shows that, at the end of an oscillation, most of the particles recover almost the initial position 
%, being confined in a single recirculating cell, 
%%they were at the beginning of the oscillation, but 
except a few particles which can escape a ripple %(and the recirculating cell) 
and reach the neighbouring one. 
%mu
%In fact, ripples do not drift during the wave cycles presently simulated.
%%
%%    Figure 19
\begin{figure}[ht]
\begin{picture}(0,120)(0,0)
  \ifnineteen
  \put(-8,-5){
    \includegraphics[width=.5\linewidth,trim=0cm 0cm 0cm 0cm,clip]{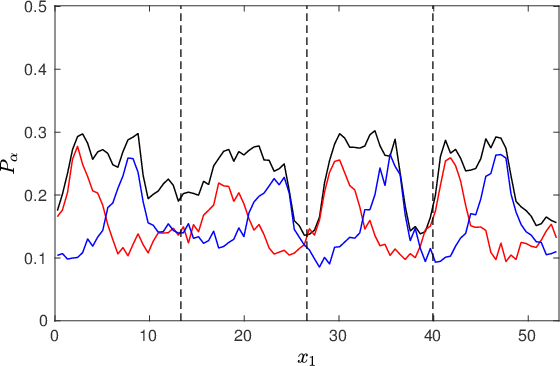}
  }
  \put(192,-5){
    \includegraphics[width=.5\linewidth,trim=0cm 0cm 0cm 0cm,clip]{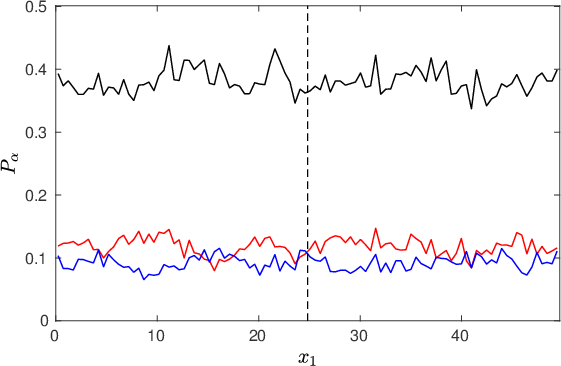}
  }
  \put(-10,114){\small $(a)$}
  \put(191,114){\small $(b)$}
  \put(16,78){\scriptsize $part.excur.$}
  \put(16,14){\color{blue!}\scriptsize $drag(-)$}
  \put(26,26){\color{red!}\scriptsize $drag(+)$}
  \put(18,20){\color{blue!}\line(0,1){11}}
  \put(220,80){\scriptsize $part.excur.$}
  \put(230,18){\color{blue!}\scriptsize $drag(-)$}
  \put(220,45){\color{red!}\scriptsize $drag(+)$}
  \fi
\end{picture}
\caption{%
Probability that top-layer particles located at the instants $t=\tin{}$ (for the last $5$ oscillation periods of each run) in a certain position $x_1$ experience semi-excursion $\alpha\equiv\vert\pse^*\vert>0.5~\wln^*$ (black lines) or time-maximum (over each half-period) drag force $\alpha\equiv\max_T\sav{\pdr^*}>+0.1~\prf^*$ (red lines) or $\alpha\equiv\min_T\sav{\pdr^*}<-0.1~\prf^*$ (blue lines). %
\mmt{Dashed} lines are equispaced by $\wln^*/\del^*$. %
The reference drag is defined as $\prf^*=\frac{1}{2}\densf^*\U^*\om^*\del^{*\,3}$. %
Probabilities are computed over the last periods of run~$1$ (panel$(a)$) and run~$2$ (panel$(b)$). %
}%
\label{fig19}
\end{figure}
\mmr{%
Figure~\ref{fig19}a shows the probability that the semi-excursion and the time-maximum (time-minimum) drag force acting on the top-layer particles selected at the instants $\tin{j}$, $j=36,\,38,\ldots,\,41$, of run~$1$ and located within $[x_1-D,\,x_1+D]$, exceed the threshold values $0.5~\wln^*$ and $+0.1~\prf^*$ ($-0.1~\prf^*$), respectively, with $\prf^*=\frac{1}{2}\densf^*\U^*\om^*\del^{*\,3}$. %
}{%
Figure~\ref{fig19}a shows the probability that the semi-excursion, the time-maximum drag force and time-minimum drag force acting on the top-layer particles selected at the instants $\tin{j}$, $j=36,\,38,\ldots,\,41$, of run~$1$ and located within $[x_1-D,\,x_1+D]$, exceed the threshold values $0.5~\wln^*$, $+0.1~\prf^*$ and $-0.1~\prf^*$, respectively, with $\prf^*=\frac{1}{2}\densf^*\U^*\om^*\del^{*\,3}$. %
}%
Similarly, figure~\ref{fig19}b refers to the 
interval between the $\tin{54}$ and $\tin{59}$, 
%half-periods of run~$2$ starting from $\tin{j}$, $j=54,\,55,\ldots,\,58$, 
which is in the final part of simulation~$2$. % presently available
For run~$1$, the values of the particle semi-excursion range between $0$ and about $0.7~\wln^*$. %
Figure~\ref{fig19}a %
\mmr{evinces}{illustrates} that the probability to observe large values of $\pse^*$ 
increases in the vicinity of the 
\mmr{crest of ripples}{ripple crests} %
while it %
\mmr{nearly halves}{is approximately halved} %
in the troughs. %
In fact, the drag force acting on crest particles is significantly larger %
\mmr{than other particles}{than the drag force acting on other particles}, %
which causes %
\mmr{them}{crest particles} %
to move longer (and farther) in the flow direction. %
Crest particles, at time $t=\tin{j}$, are not aligned along the center line between the ripple troughs
where the streamlines of recirculating cells converge (cf. figure~\ref{fig17}), %
\mmr{whereas}{instead}, %
as described in \S~\ref{ss1}, they are mostly piled on the side of each ripple opposite to the flow direction. %
Therefore, the probability curves related to the drag in figure~\ref{fig19}a are 
asymmetric with respect to the the center line of ripples. %
Another consequence of the asymmetric shape of ripple profile is that 
only a small amount of crest particles reach the neighbouring ripple 
during a half-period, although visualisations show that
several crest particles display values of $\pse^*$ larger than $0.5~\wln^*$, % 
%however only a few of them can reach the neighbouring ripple because, 
%as a consequence of the asymmetric shape of ripple profile, 
because at $t=\tin{j}$ most of them are located farther than $0.5~\wln^*$ from the downstream boundary 
between adjacent ripples. %
As a result, we observe the %
\mmr{crest of ripples}{ripple crests} %
moving to and fro over the span of $\wln^*$. %
\mmr{A s}{S}imilar dynamics can not be detected for run~$2$ by visual inspection of 
figure~\ref{fig19}b. %
%\mmr{}{% 
In this case, %
the variability of the drag force acting on top-layer particles is 
%much less pronounced than that observed for 
not as pronounced as in 
run~$1$ and, as shown in the following, %
\mmr{reflects on}{results in} %
the slower evolution of ripples. %
%
%the probability of drag force in the streamwise direction is more uniform than in run~$1$
%
%Consequently, 
Also the semi-excursion of top-layer particles is almost independent of 
the streamwise coordinate %even though it 
and 
%while it 
exhibits large values %than in case~$1$ 
because 
the average (viscous) drag acts uniformly on the 
top-layer particles and is relatively 
%viscous forces are 
strong. %er than in run~$1$ 
\mmr{note that t}{T}he drag coefficient for an isolated particle of run~$2$, 
i.e. the drag force normalized by the reference quantity 
$\frac{1}{2}\densf^*\U^{*\,2}\ds^{*\,2} $, is approximately two times larger than in run~$1$). %
%bed surface. %
%particles experience strong viscous forces. %
%in run~$2$ viscous forces play a major role in the transport of sediments. %
%}%
%
This is shown more clearly in section~\ref{ss3} where the sediment flux is related to the 
shear stress acting on the bed surface. %
%
%}% \mmr
%
% The average motion of particles during consecutive half-periods is spatially periodic of amplitude equal to the particle semi-excursion, thus ripples do not drift during the wave cycles presently simulated. 
%
%However, the values of particle semi-excursion show a rich variability during each oscillation period.
%

%\mmr{}{%
In order to understand why the formation of ripples in run~$1$ occurs significantly earlier than in run~$2$, t%
%}%
%
%Therefore, describing the motion of top-layer particles can help us to understand the origin of ripples. 
%
%T
hree quantities are presently considered 
\mmr{and are collected}{} %
for each top-layer particle throughout the simulations: the particle semi-excursion, the particle velocity and the drag force. 
Results are first shown %
\mmt{%
  in the following %
}%
for the last simulated half-period, where the differences between the motion of crest particles and of other top-layer particles are pronounced. 
The probability density function (pdf) of $\pse$ was computed for the top-layer particles of both run~$1$ and run~$2$, % shown in figure~\ref{fig10}, exhibits a mode value 
which shows that approximately $88\%$ of top-layer particles set into motion stop within a distance equal to $4~d^*$ from their (previous) rest position (see figure~\ref{fig10}a). 
\iften
%%
%%    Figure 10
\begin{figure}[ht]
\begin{picture}(0,125)(0,0)
  \put(4,10){
  \put(-4,0){
    \includegraphics[trim=0cm 0cm 0cm 0cm, clip, height=.292\textwidth]{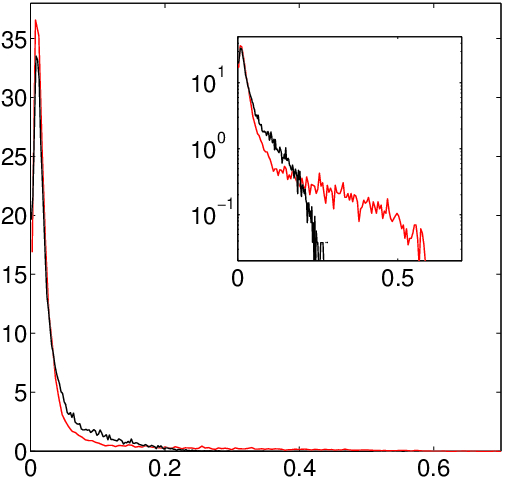}
  }
  \put(50,-8){\footnotesize $\pse^*/\wln^*$}
  \put(-14,44){\rotatebox{90}{\footnotesize $pdf^*\cdot\wln^*$}}
  \put(-16,102){\small $(a)$}
  \put(70,38){\scriptsize $\pse^*/\wln^*$}
  \put(35,62){\rotatebox{90}{\scriptsize $pdf^*\cdot\wln^*$}}
%  \put(64,98){\scriptsize \textit{crest part.}}
  \put(130,0){
    \includegraphics[trim=0cm 0cm 0cm 0cm, clip, height=.3\textwidth]{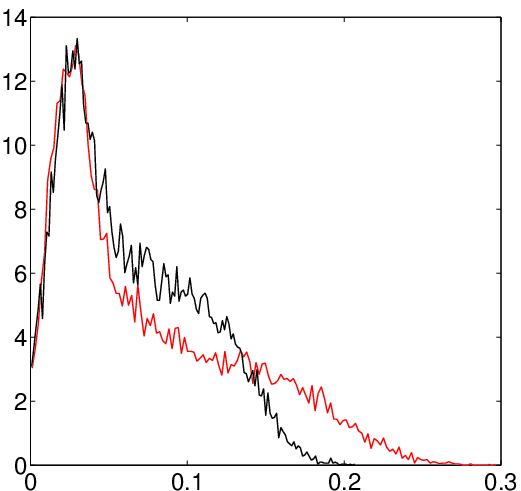}
  }
  \put(168,-8){\footnotesize $\sav{\vert\pdr^*\vert}_{\TT/2}/\prf^*$}
  \put(122,40){\rotatebox{90}{\footnotesize $pdf^*\cdot\prf^*$}}
  \put(119,102){\small $(b)$}
%  \put(170,100){\scriptsize \textit{top-layer part.}}
  \put(265,0){
    \includegraphics[trim=0cm 0cm 0cm 0cm, clip, height=.3\textwidth]{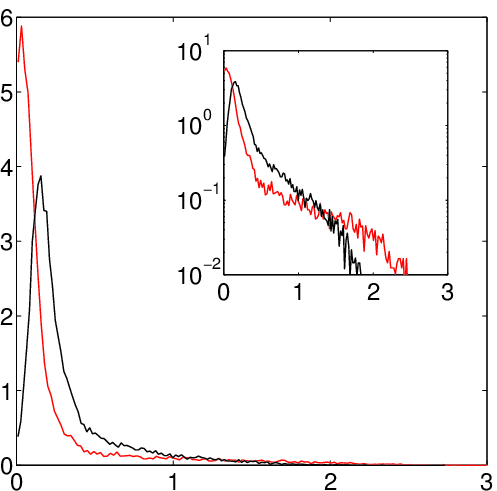}
  }
  \put(298,-8){\footnotesize $\sav{\vert\pvel^*\vert}_{\TT/2}/\om^*\del^*$}
  \put(256,40){\rotatebox{90}{\footnotesize $pdf^*\cdot\om^*\del^*$}}
  \put(254,102){\small $(c)$}
  \put(318,38){\scriptsize $\sav{\vert\pvel^*\vert}_{\TT/2}/\om^*\del^*$}
  \put(301,58){\rotatebox{90}{\scriptsize $pdf^*\cdot\om^*\del^*$}}
%  \put(302,98){\scriptsize \textit{top-layer part.}}
  }
\end{picture}
\caption{
Probability density functions of particle semi-excursion $(a)$, drag $(b)$ and velocity $(c)$ of \textit{top-layer particles} for the $41$st and $58$th half-periods of run~$1$ (red lines) and run~$2$ (black lines), respectively. Drag force is normalised by $\prf^*=\frac{1}{2}\densf^*\U^*\om^*\del^{*\,3}$. In the insets, the respective quantities are plotted in semi-logarithmic scale. 
}
\label{fig10}
\end{figure}
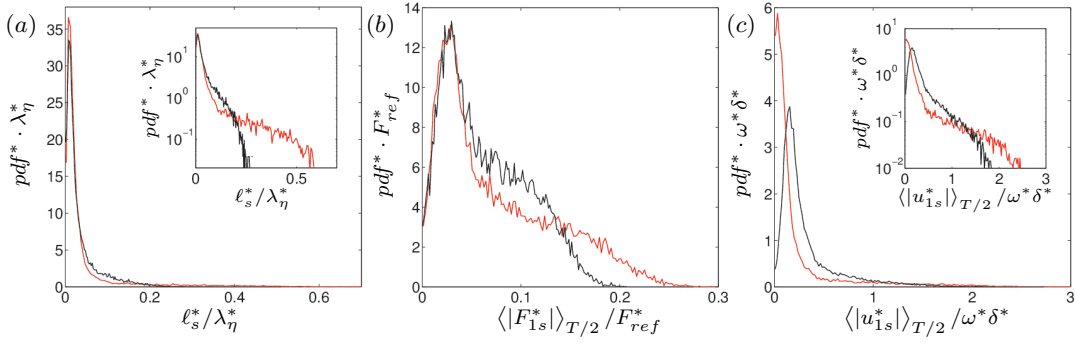
%%
%%    Figure 10a
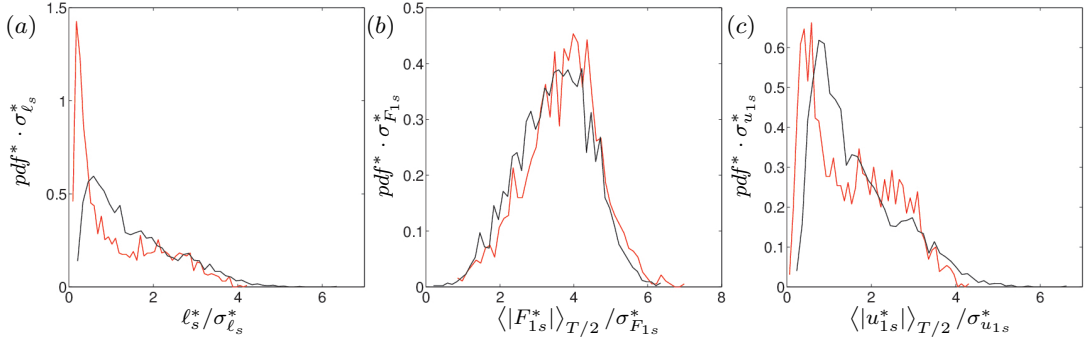
\begin{figure}[ht]
\begin{picture}(0,125)(0,0)
  \put(4,10){
  \put(-4,0){
    \includegraphics[trim=0cm 0cm 0cm 0cm, clip, height=.3\textwidth]{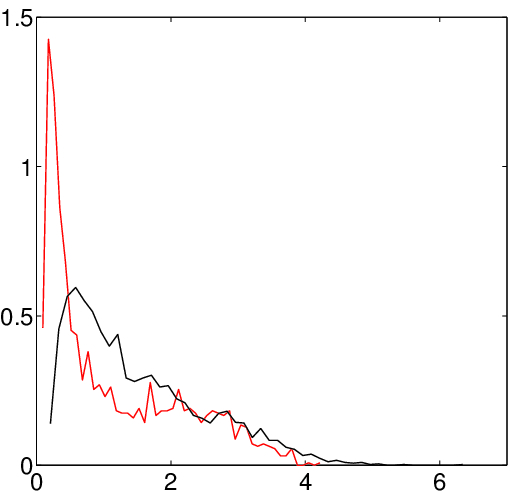}
  }
  \put(50,-8){\footnotesize $\pse^*/\sigma_{\pse}^*$}
  \put(-14,44){\rotatebox{90}{\footnotesize $pdf^*\cdot\sigma_{\pse}^*$}}
  \put(-16,102){\small $(a)$}
%  \put(64,98){\scriptsize \textit{crest part.}}
  \put(130,0){
    \includegraphics[trim=0cm 0cm 0cm 0cm, clip, height=.3\textwidth]{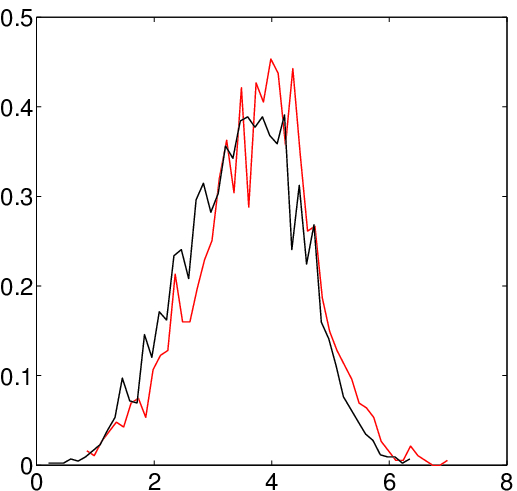}
  }
  \put(168,-8){\footnotesize $\sav{\vert\pdr^*\vert}_{\TT/2}/\sigma_{\pdr}^*$}
  \put(122,40){\rotatebox{90}{\footnotesize $pdf^*\cdot\sigma_{\pdr}^*$}}
  \put(119,102){\small $(b)$}
%  \put(198,98){\scriptsize \textit{crest part.}}
  \put(265,0){
    \includegraphics[trim=0cm 0cm 0cm 0cm, clip, height=.3\textwidth]{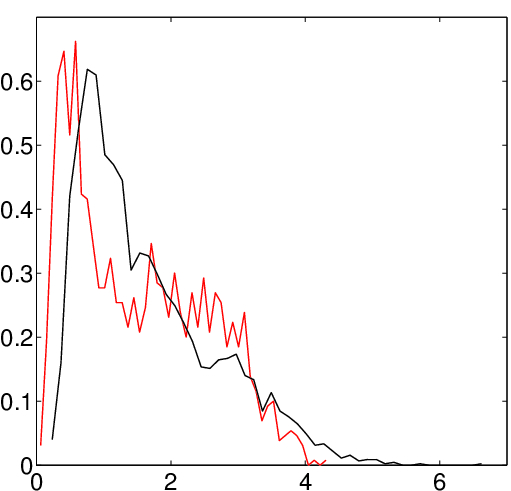}
  }
  \put(300,-8){\footnotesize $\sav{\vert\pvel^*\vert}_{\TT/2}/\sigma_{\pvel}^*$}
  \put(256,40){\rotatebox{90}{\footnotesize $pdf^*\cdot\sigma_{\pvel}^*$}}
  \put(254,102){\small $(c)$}
%  \put(330,98){\scriptsize \textit{crest part.}}
  }
\end{picture}
\caption{
Probability density functions of particle semi-excursion $(a)$, drag $(b)$ and velocity $(c)$ of \textit{crest particles} for the $41$st and $58$th half-periods of run~$1$ (red lines) and run~$2$ (black lines), respectively. Drag force is normalised by $\prf^*=\frac{1}{2}\densf^*\U^*\om^*\del^{*\,3}$. Quantities are normalised by the standard deviation of each sample. %estimated from the respective samples. 
}
\label{fig10a}
\end{figure}
\fi
Among these sluggish particles there are also crest particles that, however, predominately exhibit 
large mobility, in particular for run~$1$.
\mmc{Indeed,}\mmt{In fact,} the core of the pdfs of $\pse^*$, normalised by $\wln^*$, is found nearly coincident between run~$1$ and run~$2$ ($d^*$, $\del^*$ and $\fse^*$ are found to %
\mmr{be not}{not be} %
relevant scales of the pdf core), %scale fairly, 
while the tail of the curves, which is representative of the most mobile particles, deviates because crest particles behave differently in the two simulations and differently from the other top-layer particles. %
Such behaviour of crest particles reflects also on the particle velocity and drag, as can be
understood from figures~\ref{fig10}b,c. %(big panels). 
However, by restricting the sample to crest particles and scaling the quantities presently considered by their standard deviation, a fair matching of the pdfs can be obtained, as shown in figure~\ref{fig10a}. 
This strategy is not relevant for non-crest particles. 
The existence of two separated scales suggests that (at least) two types of particle motion coexist: a ``regular motion'' dominated by viscous forces (slow particles) and an ``erratic motion'' affected by particle-particle interactions that manifest themselves in random fluctuations of particle forces (crest particles): %
\mmc{Indeed,} about $50\%$ of crest particles of run~$1$ show a wide range of values of $\pse^*$ (between $0.10~\wln^*$ and $0.45~\wln^*$) with a nearly constant distribution of probability. %
Instead, the values of $\pse^*$ for run~$2$ are more accumulated around the mean value ($0.08~\wln^*$) 
than in the other simulation. %
This is also emphasized by the ratio between the standard deviation, $\sigma_{\pse}^*$, and the mean value, $\sav{\pse^*}_s$, being equal to $0.88$ and $0.67$ for runs~$1$ and $2$\mmt{, respectively}. %
The fact that $\wln^*$ is not a relevant scale for the semi-excursion of crest particles 
%reflects on
appears from %
the values of the statistics shown in table~\ref{tab5}. %
The standard deviation $\sigma_{\pse}^*$ computed for run~$2$ appears significantly smaller than that of run~$1$ when normalised by $\wln^*$%
\mmr{, but}{. However,} the %
\mmr{values for both the runs range}{value for each run is} %
about $0.03$ %
\mmr{if they are}{when} %
normalised by $\fse^*$, which is actually a relevant scale for the semi-excursion of crest-particles. 
The values of $\sav{\pse^*}_s$ for crest particles are equal to $3.6\cdot 10^{-2}\fse^*$ and $5.7\cdot 10^{-2}\fse^*$ %
\mmt{%
for runs $1$ and $2$, respectively, %
}%
which are smaller but of the same order (approximately half) of the value obtained by \citet{mazzuoli2016a} at the end of %
\mmr{the}{} %
test nr.~$6$. %
\mmc{Indeed,}\mmt{Actually,} in the present case there are factors that contribute to increase the friction between sediments, among them the bed surface is not 
macroscopically 
flat as in the cases investigated by \citet{mazzuoli2016a} and the number of moving particles (and, consequently, of collisions) is much larger. %
The maximum value of particle excursion, $\max_s\pse^*$, for run~$1$ is approximately equal to $0.15~\fse^*$ 
and is comparable with those computed from one of the experiments of \citet{rousseaux2004} 
($\Rdel\simeq\Rd=135$, $\s=2.5$) which fell in the range $[0.15,\,0.25]~\fse^*$. %
%\mut{COMPARISON WITH \citet{blondeaux1990} AND ROUSSEAUX (2nd MARCH GETS BACK FROM VACATION)\\
%Rousseaux ($\Rdel\simeq\Rd=135$, $\s=2.5$): $\max_s\pse^*=0.15\div 0.25\fse^*$}

%%
%%*** Table 5 *******************************************************
\begin{table}[ht]
	\begin{center}
		\begin{tabular}{l c c c c c c c}
		\hline\vspace*{.55cm}
		\multirow{2}{1.cm}{run} & 
		\multirow{2}{1.cm}{$\dfrac{\wln^*}{\fse^*}$} &
		\multirow{2}{1.cm}{$\dfrac{\sav{\pse^*}_s}{\wln^*}$} &
		\multirow{2}{1.4cm}{$\dfrac{\sav{\pdr^*}_{\TT/2,s}}{\prf^*}$} & 
		\multirow{2}{1.6cm}{$\dfrac{\sav{\vert\pvel^*\vert}_{\TT/2,s}}{\om^*\del^*}$} &
		\multirow{2}{1.1cm}{$\dfrac{\sigma_{\pse}^*}{\wln^*}$} &
		\multirow{2}{1.cm}{$\dfrac{\sigma_{\pdr}^*}{\prf^*}$} & 
		\multirow{2}{1.cm}{$\dfrac{\sigma_{\pvel}^*}{\om^*\del^*}$} \\ 
		\hline
        $1$ & $0.208$ & $0.173$ & $0.156$ & $0.953$ & $0.152$ & $0.042$ & $0.626$ \\
        $2$ & $0.693$ & $0.082$ & $0.111$ & $0.677$ & $0.055$ & $0.032$ & $0.410$ \\
		\hline
		\end{tabular}
	\end{center}
	\caption{\small Statistics of crest particles.}
	\label{tab5}
\end{table}
The discrepancies between the pdfs of the two simulations are not strictly associated with the presence 
of ripples, as one could be tempted to presume, because the same differences were present since the initial wave cycles when bedforms were not \mmt{yet} developed. 
Instead, %it is reasonable to attribute the differences of particle motion to 
they can be attributed to the different values of 
the Keulegan-Carpenter number, $\Kc$. %
%
%\mmr{}{%
In fact, large values of $\pse$ are associated with large values of $\fse$. %, and of the drag force, as shown in figure~\ref{fig10}b. 
Moreover, the contribution to the average drag force acting on the particles due to the 
presence of recirculation cells is smaller in run~$2$ than in run~$1$ which leads to more 
homogeneous distribution of drag over the bed surface. % and to slower formation of ripples. %
%}%
%
Note that most frequently (in the sense of probability) top-layer particles exhibit a creeping velocity, in particular in the case of run~$1$ which shows a wider gap between very slow and fast moving particles than run~$2$ (cf. figure~\ref{fig10}c). 
%
%\mmr{}{%
\mmr{Indeed
%,}\mmt{Thus, it can be deduced that,} 
analogously}{Analogous} to the process of segregation for poly-dispersed particulate flows, which occurs because of particle inertia when sediment particles differ in size and/or density, here the growth of bed surface perturbations is promoted by the non-uniform distribution of drag over top-layer particles (due to the steady streaming) and is faster if the non-homogeneity is more pronounced.%
%}%
%Indeed, it is reasonable that the wider is this gap the faster is the growth of bed surface perturbations. %
%

\begin{figure}[ht]
\begin{picture}(0,124)(0,0)
  \ifeleven
  \put(3,5){
  \put(-7,0){
    \includegraphics[trim=0cm 0cm 0cm 0cm, clip, height=.3\textwidth]{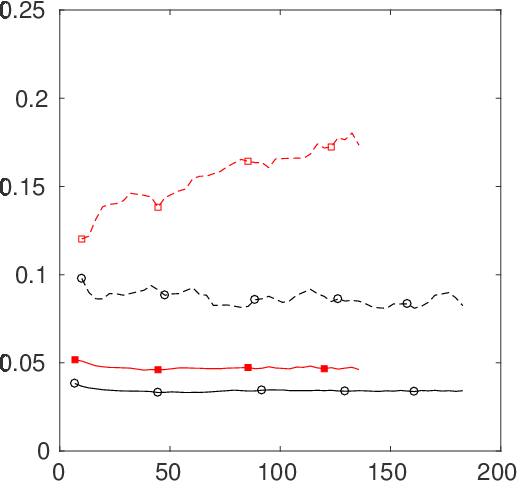}
%    \put(-10,35){$U_e$}
  }
  \put(55,-8){\footnotesize $\omega^*t^*$}
  \put(-14,44){\rotatebox{90}{\footnotesize $\sav{\pse^*}_s/\wln^*$}}
  \put(-16,105){\small $(a)$}
  }
  \put(-104,4){
  \put(237,0){
    \includegraphics[trim=0cm 0cm 0cm 0cm, clip, height=.3\textwidth]{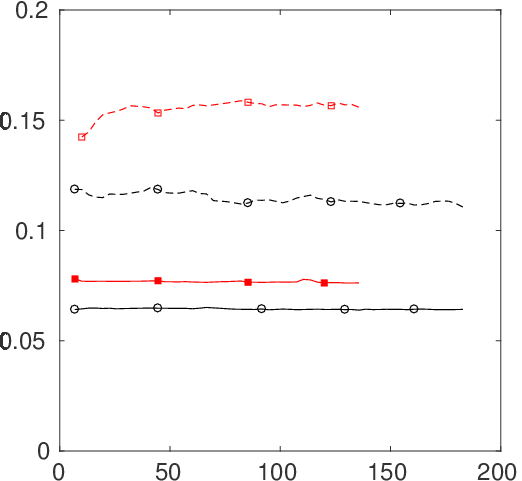}
  }
  \put(299,-8){\footnotesize $\omega^*t^*$}
  \put(228,28){\rotatebox{90}{\footnotesize $\sav{\pdr^*}_{s,\TT/2}/\prf^*$}}
  \put(228,105){\small $(b)$}  
  }
  \put(150,4){
  \put(122,0){
    \includegraphics[trim=0cm 0cm 0cm 0cm, clip, height=.3\textwidth]{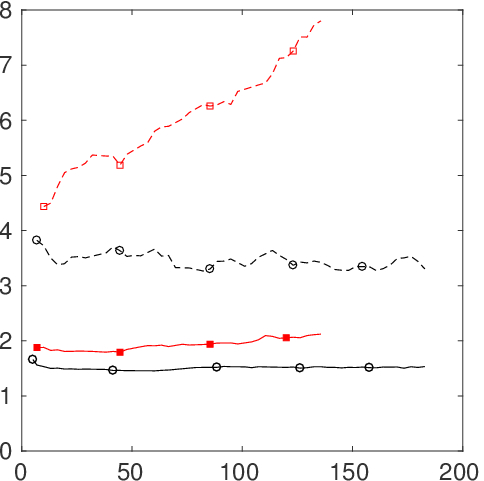}
  }
  \put(175,-8){\footnotesize $\omega^*t^*$}
  \put(112,12){\rotatebox{90}{\footnotesize $\sav{\max_{\TT/2}\vert\pvel^*\vert}_s/\om^*\del^*$}}
  \put(110,105){\small $(c)$}
  }
  \fi
\end{picture}
\caption{Statistics of the motion of top-layer (solid lines) and crest (\mmt{dashed} lines) particles: $(a)$ mean particle semi-excursion, $(b)$ maximum velocity, $(c)$ mean time-maximum drag.
The values are computed for each half-period of simulation~$1$ (red lines/squares) and simulation~$2$ (black lines/circles).
Particle semi-excursion for run~$1$ increases approximately with a linear trend: similar trend was observed by \citet{mazzuoli2016a} in their tests no. 2,3,4.
}
\label{fig11}
\end{figure}
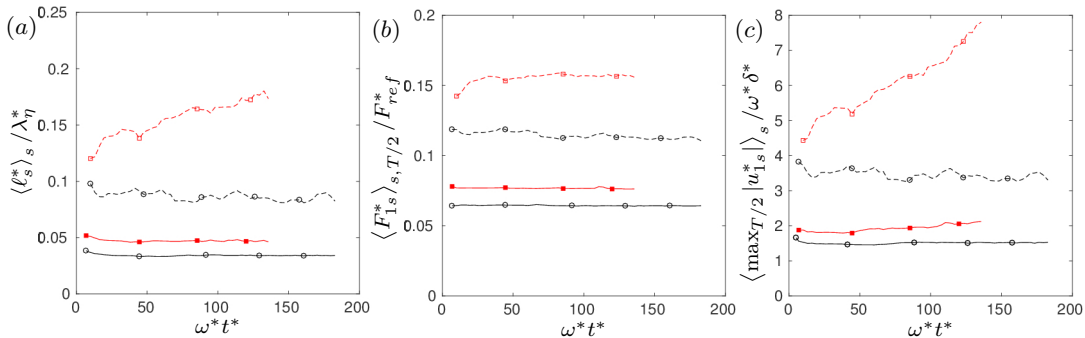

%
In principle, the mechanism at the origin of the patterns of spheres on 
the surface of a movable bed is similar to that observed by \citet{mazzuoli2016a} 
for beads rolling on a rough plane bottom (tests~$4$ and $6$). %
In %
\mmr{the}{} %
test no.~$6$ of \citet{mazzuoli2016a}, several movable beads were initially aligned in the 
direction of flow oscillations  and rapidly spread laterally until, in a 
few oscillations, they were randomly scattered over the whole bottom. %
The latter is approximately the initial configuration of the present simulations. %
The evolution of the values of $\pse$, $\pdr$ and $\pvel$, averaged over top-layer or crest particles and over each half-cycle, are shown in figure~\ref{fig11}. 
%
%%
%%    Figure 11
The values of the considered quantities for top-layer particles are approximately constant throughout simulation~$2$, except a short initial transient, while a slight monotonic decrease of the three quantities can be noted %
\mmr{relatively}{relative} %
to crest particles.
During the transient, both the semi-excursion and the velocity of top-layer particles decrease 
because the spheres attain a closely packed configuration.
%
%\mmr{}{%
Conversely, crest particles of run~$1$ manifest increasing mobility since the beginning of 
the simulation and show an approximately linear growth of the average $\pse$ and $\pvel$, while drag seems to %
\mmr{reach asymptotically}{asymptotically reach} %
a constant value after the initial transient. %
%}%
%
\mmr{%
The former phenomenon 
}{%
The increase of mobility of crest particles %
}%
is due to the emergence of ripples which push crest particles towards regions of the boundary layer characterized by higher velocity. 
The effect of the exponential growth of the %
\mmr{amplitude of ripples}{ripple amplitudes} %
is partly opposed by that of increasing inter-particle collisions. %
Since the inertia of particles is relatively small in both runs~$1$ and $2$ (as indicated by the large values of $\Kc$), the drag force reaches a limit %
\mmr{value}{} %
as it balances the bed friction. %
Thus, the %
\mmr{particle relative velocity}{relative particle velocity}, on which the viscous drag depends, 
remains constant while the particle \mmr{absolute velocity increases}{absolute particle velocity}. %
In fact, as %
\mmr{it}{} %
will be clarified in the following section, the viscous drag dominates the other force 
contributions in the phases when the bed shear stress is \mmt{at a} maximum. %

\iftabs34
%%
%%*** Table 3 *******************************************************
\begin{table}[ht]
	\begin{center}
		\begin{tabular}{l l r c c c}
		\hline
		\multirow{1}{*}{} & 
		\multirow{1}{*}{run} & 
        \multirow{1}{*}{$\#$half-period} & 
		\multirow{1}{*}{mode} &
		\multirow{1}{*}{mean} & 
		\multirow{1}{*}{max} \\ 
		\hline
		$\pse^*/\fse^*$  & $1$ & $2$  & $1.4\cdot10^{-3}$  & $2.5\cdot10^{-2}$ & $1.4\cdot10^{-1}$ \\
        $\pvel^*/\U^*$   & $1$ & $2$  & $1.5\cdot10^{-2}$  & $6.8\cdot10^{-2}$ & $3.3\cdot10^{-1}$ \\
        $\pdr^*/\prf^*$  & $1$ & $2$  & $1.5\cdot10^{-1}$  & $1.4\cdot10^{-1}$ & $2.9\cdot10^{-1}$ \\
		\hline
		$\pse^*/\fse^*$  & $1$ & $41$ & $1.6\cdot10^{-3}$  & $3.6\cdot10^{-2}$ & $1.6\cdot10^{-1}$ \\
        $\pvel^*/\U^*$   & $1$ & $41$ & $3.9\cdot10^{-2}$  & $3.9\cdot10^{-1}$ & $1.2\cdot10^{-1}$ \\
        $\pdr^*/\prf^*$  & $1$ & $41$ & $1.7\cdot10^{-1}$  & $1.6\cdot10^{-1}$ & $3.0\cdot10^{-1}$ \\
		\hline
		$\pse^*/\fse^*$  & $2$ & $2$  & $6.0\cdot10^{-3}$  & $7.1\cdot10^{-2}$ & $2.3\cdot10^{-1}$ \\
        $\pvel^*/\U^*$   & $2$ & $2$  & $3.2\cdot10^{-2}$  & $1.1\cdot10^{-1}$ & $3.1\cdot10^{-1}$ \\
        $\pdr^*/\prf^*$  & $2$ & $2$  & $1.2\cdot10^{-1}$  & $1.2\cdot10^{-1}$ & $2.0\cdot10^{-1}$ \\
		\hline
		$\pse^*/\fse^*$  & $2$ & $58$ & $5.4\cdot10^{-3}$  & $5.7\cdot10^{-2}$ & $2.4\cdot10^{-1}$ \\
        $\pvel^*/\U^*$   & $2$ & $58$ & $3.6\cdot10^{-2}$  & $9.2\cdot10^{-2}$ & $3.6\cdot10^{-1}$ \\
        $\pdr^*/\prf^*$  & $2$ & $58$ & $1.1\cdot10^{-1}$  & $1.1\cdot10^{-1}$ & $2.0\cdot10^{-1}$ \\
		\hline
		\end{tabular}
	\end{center}
	\caption{\small Statistics of the particle motion (crest particles).}
	\label{tab3}
\end{table}
%%
%%*** Table 4 *******************************************************
\begin{table}
	\begin{center}[ht]
		\begin{tabular}{l l r c c c}
		\hline
		\multirow{1}{*}{} & 
		\multirow{1}{*}{run} & 
        \multirow{1}{*}{$\#$half-period} & 
		\multirow{1}{*}{mode} &
		\multirow{1}{*}{mean} & 
		\multirow{1}{*}{max} \\ 
		\hline
		$\pse^*/\fse^*$  & $1$ & $2$  & $0$                & $1.0\cdot10^{-2}$ & $1.4\cdot10^{-1}$ \\
        $\pvel^*/\U^*$   & $1$ & $2$  & $5.3\cdot10^{-3}$  & $2.9\cdot10^{-2}$ & $3.3\cdot10^{-1}$ \\
        $\pdr^*/\prf^*$  & $1$ & $2$  & $2.4\cdot10^{-2}$  & $7.7\cdot10^{-2}$ & $2.9\cdot10^{-1}$ \\
		\hline
		$\pse^*/\fse^*$  & $1$ & $41$ & $0$                & $9.6\cdot10^{-3}$ & $1.5\cdot10^{-1}$ \\
        $\pvel^*/\U^*$   & $1$ & $41$ & $6.6\cdot10^{-3}$  & $3.3\cdot10^{-2}$ & $4.0\cdot10^{-1}$ \\
        $\pdr^*/\prf^*$  & $1$ & $41$ & $2.8\cdot10^{-2}$  & $7.6\cdot10^{-2}$ & $3.0\cdot10^{-1}$ \\
		\hline
		$\pse^*/\fse^*$  & $2$ & $2$  & $0$                & $2.7\cdot10^{-2}$ & $2.3\cdot10^{-1}$ \\
        $\pvel^*/\U^*$   & $2$ & $2$  & $1.5\cdot10^{-2}$  & $4.4\cdot10^{-2}$ & $3.4\cdot10^{-1}$ \\
        $\pdr^*/\prf^*$  & $2$ & $2$  & $2.9\cdot10^{-2}$  & $6.4\cdot10^{-2}$ & $2.1\cdot10^{-1}$ \\
		\hline
		$\pse^*/\fse^*$  & $2$ & $58$ & $0$                & $2.4\cdot10^{-2}$ & $2.4\cdot10^{-1}$ \\
        $\pvel^*/\U^*$   & $2$ & $58$ & $1.1\cdot10^{-2}$  & $4.3\cdot10^{-2}$ & $3.6\cdot10^{-1}$ \\
        $\pdr^*/\prf^*$  & $2$ & $58$ & $3.0\cdot10^{-2}$  & $6.4\cdot10^{-2}$ & $2.1\cdot10^{-1}$ \\
		\hline
		\end{tabular}
	\end{center}
	\caption{\small Statistics of the particle motion (top-layer particles).}
	\label{tab4}
\end{table}
\fi

\subsection{Bed shear stress, incipient particle motion and sediment flow rate}\label{ss3}
%%
%\subsection{Sediment flux and condition of incipient sediment transport}\label{ss3}
%\textit{Identify phases of (surficial) sediment transport, distinguish added mass (read "pressure gradient") effects from drag effects on sediment, comparison with empirical relationship established for steady flows}\\

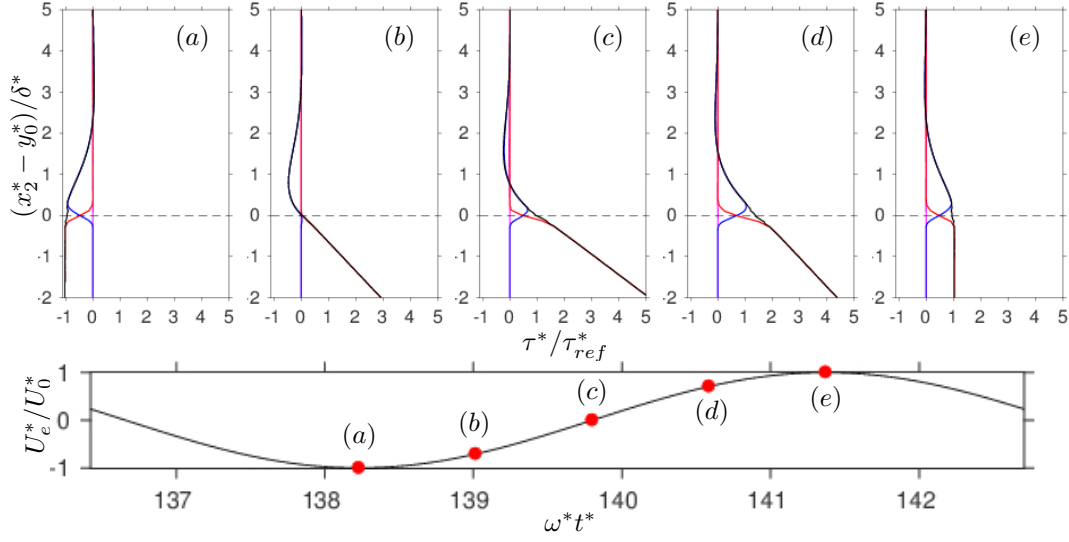
\begin{figure}%[ht]
%  \centering
%    \centerline{(\textit{a})}
  \begin{picture}(0,200)(0,0)
  \iftwentytwo
  %\begin{minipage}{2ex}
  \put(0,4){
  \put(-10,110){  \rotatebox{90} {$(x_2^*-y_0^*)/\del^*$}}
  %\end{minipage}\\
%  \begin{minipage}{0.96\textwidth}
%    \centerline{(\textit{a})}
  \put(4,70){\includegraphics[width=0.19\textwidth,trim=.07cm 0 0 0,clip]
          {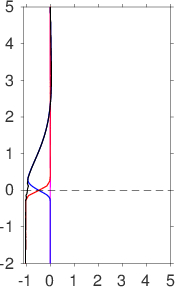}}
  \put(55,175){$(a)$}
  \put(82,70){\includegraphics[width=0.19\textwidth,trim=.07cm 0 0 0,clip]
          {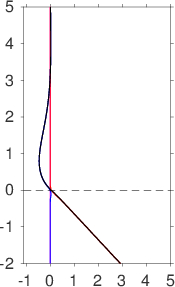}}
  \put(133,175){$(b)$}
  \put(160,70){\includegraphics[width=0.19\textwidth,trim=.07cm 0 0 0,clip]
          {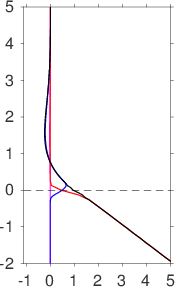}}
  \put(211,175){$(c)$}
  \put(238,70){\includegraphics[width=0.19\textwidth,trim=.07cm 0 0 0,clip]
          {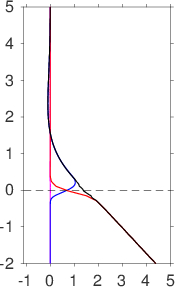}}
  \put(289,175){$(d)$}
  \put(316,70){\includegraphics[width=0.19\textwidth,trim=.07cm 0 0 0,clip]
          {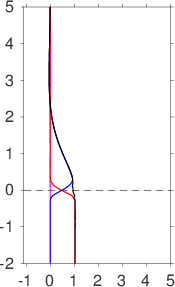}}              
  \put(367,175){$(e)$}
%    \centerline{ $\tau/\tau_{ref}$}
  \put(185,60){$\tau^*/\tau_{ref}^*$}
  }
%  \end{minipage}\\[10pt]
  %
%  \begin{minipage}{2ex}
  \put(-5,25){  \rotatebox{90} {$U_e^*/\U^*$}}
%  \end{minipage}
%   \begin{minipage}{0.96\textwidth}
%    \centerline{ $(b)$}       
  \put(5,4){  \includegraphics[width=0.96\textwidth,trim=0 0 0 0cm,clip]
                {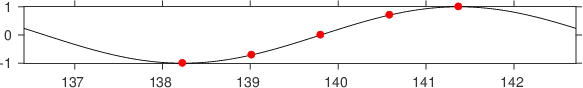}}
%             \centerline{ $\om t$}
  \put(117.5,29){$(a)$}
  \put(161,33.5){$(b)$}
  \put(205,45.5){$(c)$}
  \put(249.5,38){$(d)$}
  \put(292.5,43){$(e)$}
%  \end{minipage} 
  \put(190,-4){  $\om^* t^*$}
  \fi % \iftwentytwo
  \end{picture}
   \caption{\protect\label{fig:shear-stress-profiles-S4}%
   %(\textit{a}) 
    Sample wall-normal profile of the fluid viscous shear stress (blue line),
    Reynolds shear stress (magenta line), stress stemming from the
    fluid-particle interaction (red line) as well as the total
    shear stress $\tautot$ 
    (black line). The profiles correspond to 
    selected times which are indicated in bottom panel. %(\textit{b}).
    Data corresponds to run 1.}
\end{figure}

%
The wall-normal dependent total shear stress 
%\mmr{}{acting on the (rippled) bed }
is a sum of the fluid
shear stress $\tauf^*$ and the contribution stemming from the
fluid-particle interaction $\taup^*$, viz.
\begin{equation}
  \tautot^* = \tauf^* + \taup^*
\end{equation}
where the fluid shear stress (under a turbulent flow condition) is
comprised of the viscous and Reynolds shear stress contributions:
\begin{equation}
  \tauf^*(x_2^*) 
  = 
  \densf^*\nu^*\dfrac{\partial\sav{u_1^*}}{\partial x_2^*}(x_2^*) 
  - 
  \densf^*\sav{u_1^{\prime *} u_2^{\prime *}}(x_2^*)\;,
\end{equation}
%\mmr{}{
where the dependence on $t^*$ is omitted for the sake of clarity. %
  While $x_3$ is a homogeneous direction for the total shear stress also 
  in the presence of ripples, 
  relatively small fluctuations about $\tautot^*$ can be observed in 
  the streamwise direction when the rolling-grain ripples form. %
  Thus, $\tautot^*$ is the average total shear stress acting on the bed. %
%}%
%
It is expected that, in the present configuration, the Reynolds shear
stress has negligible contribution as the flow is essentially laminar. %
In the context of the immersed boundary method, the stress exerted by
the particles is given by
\begin{equation}
  \taup^*(x_2^*) = \densf^*\int_{x_2^*}^{L_{x_2}^*}\sav{f_1^*} {\rm d}x_2^*
\end{equation}
where $f_1^*$ is the streamwise component of the immersed boundary method 
volume forcing exerted on the fluid, transferred to the Eulerian grid 
(cf.\ \S\ref{appr}). %
In a stationary channel
flow scenario, the total shear stress varies linearly in the
wall-normal direction with a slope equal to the value of the
imposed driving pressure gradient. %
In the present OBL configuration however, as a result of the 
non-stationarity, $\tautot^*$ responds to
the pressure gradient in a complex non-linear behavior. %
Figure~\ref{fig:shear-stress-profiles-S4} shows sample wall-normal
profiles of the different contributions to the total shear stress, 
non-dimensionalised by $\tau_{ref}^*=\frac{1}{2}\densf^*\U^*\om^*\del^*$, 
at different time instants. %
As is expected, the contribution from the Reynolds 
shear stress is negligibly small across the entire wall-normal interval
which is a further indication that the flow has not separated behind %(in front) of 
the ripple crests. %
In the clear fluid region, that is, in the region which is essentially
devoid of sediment particles, only the fluid viscous shear stress
contributes to $\tautot$. %
On the other hand, deep inside the sediment bed sufficiently below the 
fluid-bed interface, $\tauf$ vanishes and $\tautot$ entirely is comprised 
of the stress exerted by the sediment particles. %
It is worth noting that, in this region, $\tautot$ exhibits a
linear variation with a slope equal to the imposed pressure gradient. %
This means that the particle shear resistance, which is proportional to
the submerged weight of sediment bed, is instantaneously in
equilibrium with the oscillating driving force (neglecting the small particle
velocities in this region). %
In between these two regions, there exists a third ``active layer'' 
region, hereafter referred to as \textit{mobile layer}, where both 
$\tauf$ and $\taup$ contribute to the total shear 
stress and where %most of 
the particle erosion-deposition processes take place. %
Although there is no clear demarcation of these regions, it is
observed that the thickness of the mobile layer varies depending on
different phases of the oscillation period. %
%==========================================================
%figure: shear stress profiles

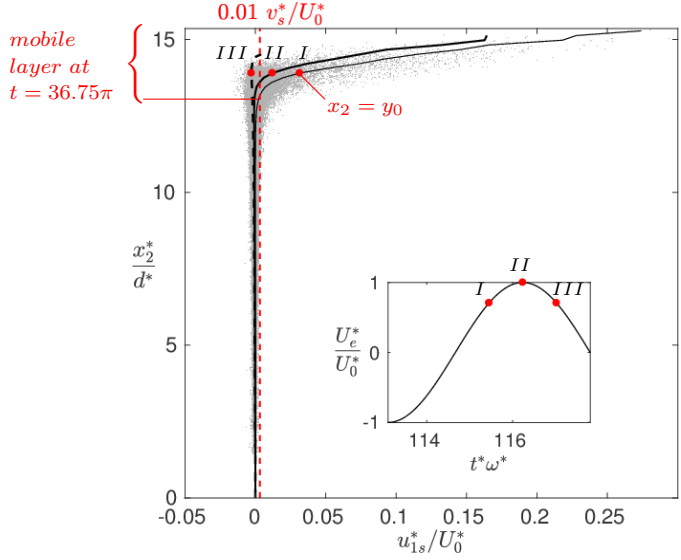
\begin{figure}%[ht]
%  \centering
%    \centerline{(\textit{a})}
  \begin{picture}(0,210)(0,0)
  \iftwentythree
    \put(81,-3){\includegraphics[width=0.532\textwidth,trim=0 0 0 0cm,clip]
                    {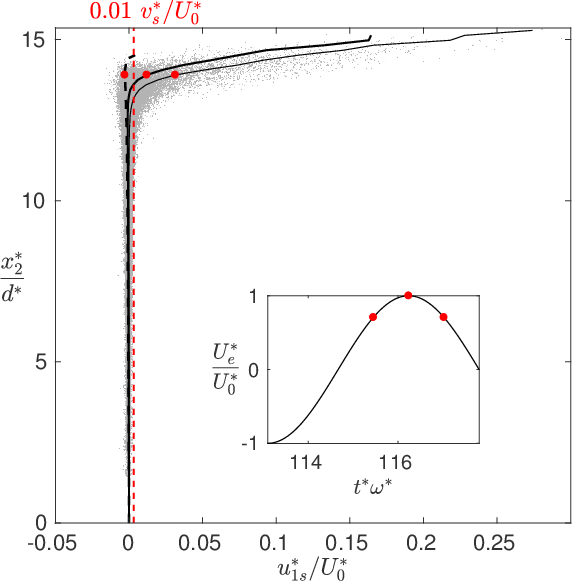}}
    \put(153,168){\color{red!}\line(-1,1){10}}
%    \put(153.5,160){\color{red!}{\footnotesize $\left. \dfrac{x_2^*}{\ds^*}\right\vert_{\svf=0.1}$}}
%    \put(154,163){\color{red!}{\footnotesize $\svf=0.1$}}
%    \put(154,163){\color{red!}{\footnotesize $mobile\,layer\,interface$}}
    \put(154,163){\color{red!}{\footnotesize $x_2=y_0$}}
    \put(209,94){\color{black!}{\scriptsize $I$}}
    \put(222,103){\color{black!}{\scriptsize $II$}}
    \put(238,94){\color{black!}{\scriptsize $III$}}
    \put(143,183){\color{black!}{\scriptsize $I$}}
    \put(130,183){\color{black!}{\scriptsize $II$}}
    \put(112,183){\color{black!}{\scriptsize $III$}}
    \put(85,168){\color{red!}\line(1,0){43}}
    \put(78,179){\color{red!}$\Bigg\{$}
    \put(35,188){\footnotesize\color{red!}$mobile$}
    \put(35,178){\footnotesize\color{red!}$layer\ at$}
    \put(35,168){\footnotesize\color{red!}$t=36.75\pi$}
  \fi % \iftwentythree
  \end{picture}
   \caption{\protect\label{fig:mobile-layer-definition}%
   %(\textit{a}) 
    In the main panel, the streamwise component of the particle velocity is plotted as a function of the wall-normal coordinate of run~$1$. Shaded by grey dots are the velocity of each particle at the instants %$t=36.38\pi$, 
    $t=36.75\pi$ (I), $t=37.00\pi$ (II) and $t=37.25\pi$ (III) (phases are indicated in the small inset), while %thin \mmt{dashed}, 
    thin solid, thick solid and \mmt{dashed} lines indicate the respective (binned) average values. %When the bed shear stress, $\tau_b$, attains the maximum value (approximately at $t=36.75\pi$), 
%It can be noted that at $t=36.38\pi$, the position of $y_0$ (red dots), where $\sav{\svf}$ exceeds $0.1$, approaches the elevation at which the particle velocity profile reaches the threshold of particle velocity equal to $0.01~\vs$.
}
\end{figure}

For modeling purposes, it is common practice to relate the
non-dimensional boundary shear stress $\tau_b^* = \tautot^*(x_2^*=y_0^*)$,
i.e.\ the Shields number
\begin{equation}
  \shields = \frac{\tau_b^*}{(\denss^*-\densf^*)g^*d^*}
\end{equation}
to the sediment flow rate. %
The value of $y_0^*$ is chosen as the distance from the %
\mmr{%
wall %
}{%
bottom %
}% 
at which the average particle 
volume fraction $\sav{\svf}$ reaches $0.1$. %
The instantaneous volumetric flow rate of
the particle phase (per unit span), $\qpmean^*$, is given by
\begin{equation}\label{eq:particle-flow-rate}
  \qpmean^*(t^*)  = \frac{\pi \ds^{*\,3}}{6\,\Lx^*\Lz^*}
  \sum_{l=1}^{N_p}\pvel^{*(l)}(t^*)\,,
\end{equation}
where $\pvel^{*(l)}(t^*)$ is the streamwise component of the %instantaneous
velocity of the $l$-th mobile particle at time $t^*$. %
%
%\mmr{}{%
%As a result of the 
Since spherical particles do not gear to each other 
and can slide more easily than sand grains, 
%discrete nature of the system, 
many particles experience non-zero
velocities, even if they are located below the bed surface. %
%}%
%
Thus, in order to exclude all particles which do not 
contribute to the shear induced particle flux, a streamwise velocity
threshold is set at $1\%$ of the gravitational velocity of particles, $\vs^*$ 
(similar results are obtained even considering 
the threshold at 
a small percentage of $\om^*\del^*$). %
The particle selected with such criterion approximately coincide with those constituting the 
mobile layer (cf. figure~\ref{fig:mobile-layer-definition}). %
%==========================================================
%figure: shear stress profiles

%Moreover, only particles with wall-normal location larger than $8D$ are
%considered. 
%\aman{Here the text from Aman}.

%==========================================================
%figure: time evolution of shields number and particle flowrate
%% \begin{figure}[h]            %
%%   \centering
%%   \begin{minipage}{2ex}
%%     \rotatebox{90} {$\shields$, $SI$}
%%   \end{minipage}
%%   \begin{minipage}{0.45\textwidth}
%%     \includegraphics[width=\linewidth]
%%                     {./figure_aman/ShearStressParticleFlowrateEvolution_S4fine.eps}
%%              \centerline{ $\mod(\om t,2\pi)$}       
%%   \end{minipage}
%%   %
%%   \begin{minipage}{0.45\textwidth}
%%     \includegraphics[width=\linewidth]
%%                     {./figure_aman/ShearStressParticleFlowrateEvolution_S5fine.eps}
%%              \centerline{ $\mod(\om t,2\pi)$}       
%%   \end{minipage}
%%    \begin{minipage}{2ex}
%%     \rotatebox{90} {$\qpmean/\qref$}
%%   \end{minipage}
%%   \caption{\protect\label{fig:shields-particle-flowrate-evolution-S4S5}%
%%     Time evolution of the Shields number \shields\ (black) as well as the
%%     normalized particle flow rate (red) for the last of approximately
%%     four periods.
%%     $(a)$ run 1 $(b)$ run 2}
%% \end{figure}      

%==========================================================
%figure: Shields number vs particle flow rate
\begin{figure}[ht]
  \centering
  %% \begin{minipage}{2ex}
  %%   \rotatebox{90} {$\qpmean/\qref$}
  %% \end{minipage}
  %% \begin{minipage}{0.45\textwidth}
  %%   \centerline{ $(a)$}    
  %%   \includegraphics[width=\linewidth]
  %%                   {./figure_aman/ShearStress_vs_ParticleFlowrate_S4S5.eps}
  %%            \centerline{ $\shields$}       
  %% \end{minipage}
  %
  \begin{minipage}{2ex}
    \rotatebox{90} {$\qquad\vert\qpmean^*\vert/(\ds^*\vs^*)$}
  \end{minipage}
  \begin{minipage}{0.6\textwidth}
    \iftwentyfour
    % \centerline{ $(b)$}   
    \includegraphics[width=\textwidth,trim=0 0 0 0,clip]
                    {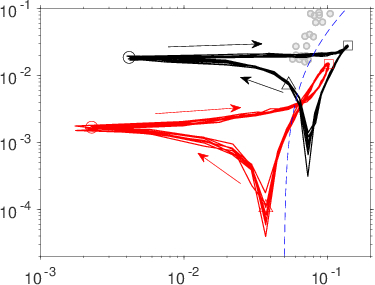}
%                    {\figdir ShearStress_vs_ParticleFlowrate_S4fine_loglog}
             \centerline{$\qquad\vert\shields\vert$}
    \fi % \iftwentyfour
  \end{minipage}
  \caption{\protect\label{fig:shields-vs-particle-flowrate-S4S5}%
    Instantaneous dimensionless particle flow rate, normalized by the inertial scaling
    $\ds^*\vs^*$, as a function of the Shields number $\shields$ during the last
    four cycles of the simulation interval.
    Run~$1$ (red line); run~$2$ (black line).
    The dashed-line represents the the Meyer-Peter \& M\"uller formula
    \citep{wong2006} for steady turbulent flow conditions
    $\qpmean = 4.93(\shields-\shields_c)^{1.6}$.
    The symbols $\bigtriangleup$, $\bigcirc$, $\Box$ indicate phases
    $t = 9/8\pi$, $t=41/32\pi$ and $t=7/4\pi$. 
    %\aman{note that the circle symbols
    %  are two flow fields before the time $t=41/32\pi$.
    %  Also, I don't understand what these phases values are, I just reproduced
    % from you Italian report.}
    %
    Gray circles indicate the experimental observations of 
    Gilbert %(1914) 
    and Meyer-Peter %(1934) 
    \citep{Nielsen1992}. %
    %\citet{gilbert1914} and \citet{meyer1934}. %
    }
\end{figure}      

Figure~\ref{fig:shields-vs-particle-flowrate-S4S5} shows the absolute value 
of the particle flow rate, normalised by $\ds^*\vs^* $, as a function of the 
absolute value of the Shields number for the last four periods of run~$1$ 
and run~$2$. %
%
%\mmr{}{
The arrows indicate the time-development along the loop %covered in the time-span of 
swept in 
a half-period. %
%
%During each half-period t
Following the loop, the particle flow rate exhibits a minimum in the early deceleration phase. %
%(i.e. when the free-stream velocity decelerates), 
Then, while the free-stream velocity is still decelerating and the Shields number decreasing, 
the particle flow rate increases under the action of the imposed driving pressure gradient. %
Subsequently, the next acceleration phase starts and the Shields number rises. %
Finally, approximately 
\mmr{%
$1/4$ of half-period after the flow reversal, %
}{%
at the phase $\frac{3}{4}\pi$ (i.e. $\pi/4$ after the flow reversal), %
}%
both the Shields number 
(i.e. the boundary shear stress) and the sediment flow rate are maximum. %
It can be noted that, 
except for two or three instants, 
%in many cases 
each value of $\vert\shields\vert$ %in the ranges of the present simulations 
correspond to two values of the dimensionless 
sediment flow rate. %
It is therefore clear, by comparing the diagrams of figure~\ref{fig:shields-vs-particle-flowrate-S4S5} 
obtained for the present runs with the experimental measurements of Gilbert %(1914) 
and Meyer-Peter %(1934) %(grey symbols) 
\citep[grey symbols, ][]{Nielsen1992} 
and with the Meyer-Peter \& M\"uller formula \citep[\mmt{dashed} line, ][]{wong2006} obtained 
for stationary channel flows, 
that the effects of the flow unsteadiness reflect strongly on the motion of particles 
and should be taken into account in the models of sediment transport. %
%it is clear that models of the sediment transport should 
%consider the effect of unsteadiness should to avoid coarse prediction errors. %
%
Indeed, during the flow reversal, which is characterised by large 
values of the forcing pressure gradient and relatively small values of the bed shear stress, the 
sediment flow rate is not negligible. %
Hence, coarse prediction errors could be avoided by relating the sediment flow 
rate to a combination of the Shields number and some dimensionless expression 
of the pressure gradient such as the instantaneous Sleath parameter \citep{foster2006,frank2015}, defined as: %
%\mmr{}{%
%
\begin{equation}
\mathcal{S} 
= 
-\dfrac{\ds^*}{\densf^*\vs^{*2}}
\dfrac{dp_f^*}{dx_1^*} 
= 
%2\mob\dfrac{\Rd}{\Rdel^2} \dfrac{dp_f}{dx_1}\:\: .
\dfrac{\mob}{\Kc} \sin{(t)}\:\: ,
\label{eq-ppres}
\end{equation}
%which, for the present simulations, is equal to $\mob/\Kc\,\sin{(t)}$. %
where the expression \eqref{pres-g} was substituted in the second equality. %to obtain the right side equality. %
%}%
%
Since the values of the Keulegan-Carpenter number, $\Kc$, are large in both the present simulations, 
the contribution of the viscous drag is expected to dominate over that induced on the spheres by 
the pressure gradient. %
Besides the direct contribution on the particle force, the pressure gradient %
\mmr{causes also}{also causes} %
the acceleration of the interstitial fluid which, due to its small inertia, responds much 
earlier than the clear fluid above the bed. %
Thus, such viscous pore flow develops and can mobilize the sediment particles 
much earlier than the bed shear stress becomes appreciable. %eventually causing their
%and contributes 
%to mobilise them earlier than the bed shear stress becomes appreciable. %
%
Even though the velocity of the particles set into motion during the flow reversal is small, 
the thickness of the mobile layer is relatively large in these phases because the pressure 
gradient acts uniformly on the entire bed. %
Finally, as the values of $\vert\shields\vert$ become large, crest particles, 
which are more exposed to the flow in the boundary layer and exhibit values 
of streamwise (particle) velocity of order $\mathcal{O}(0.1)~\U^*$, 
mostly contribute to the sediment flow rate. %
%
%}%
%

\mmt{%
Figure~\ref{fig:shields-vs-particle-flowrate-S4S5} also shows that, at corresponding 
phases, the Shields number and, therefore, the sediment flow rate normalised by 
$\ds^*\vs^*$, are larger in run~$2$ than in run~$1$. %
To understand such difference it is useful to compare the Shields number 
to that we would observe in absence of sediments, i.e. in a Stokes boundary layer:
%
%In particular, the maximum values attained by the Shields and the Sleath numbers in a Stokes boundary layer are:
\begin{equation}
\shields_{(St)} 
= 
\dfrac{\mob}{\Rdel}\left[\sin{(\om^*t^*)} 
- 
\cos{(\om^*t^*)}\right] \:\:.
%\qquad\text{and}\qquad 
%\mathcal{S} = 
\label{eqSt}
\end{equation}
\begin{figure}%[ht]
  \begin{picture}(0,135)(0,-5)
  \put(5,-5){
  \put(0,-2){
    \put(0,0){\includegraphics[width=.48\textwidth]{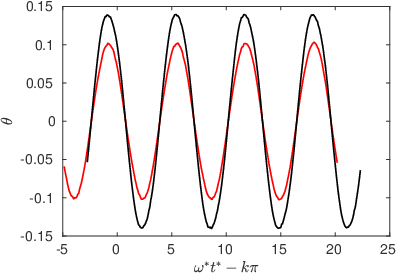}}
  }
  \put(-3,115){\small $(a)$}
  \put(190,-2){
    \put(0,0){\includegraphics[width=.48\textwidth]{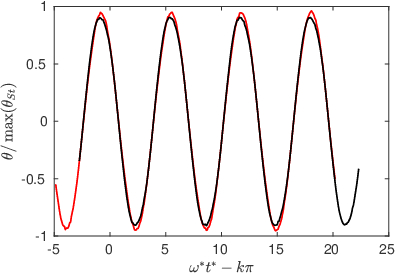}}
  }
  \put(191,115){\small $(b)$}
  }
  \end{picture}
  \label{fig:stokes}
  \caption{%
  \mmt{%
  Evolution of the Shields number for run~$1$ (red line) and run~$2$ (black line) during the last simulated periods. In panel~$(b)$ the Shields number is normalized by the maximum Shields number attained in the absence of particles, i.e. in a Stokes boundary layer. %
  }%mmt
  }%
\end{figure}
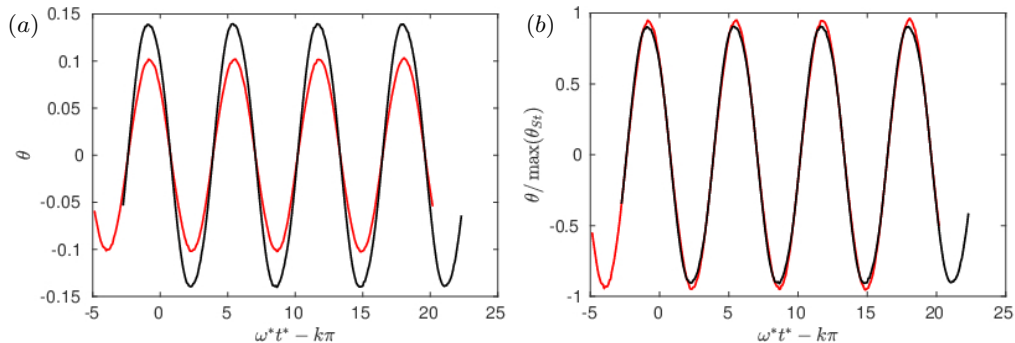
Figure~\ref{fig:stokes} shows that, scaling the Shields number by the maximum value of $\shields_{(St)}$, 
%(or, similarly, the bed shear stress normalised by $\frac{1}{2}\densf^*\U^*\om^*\del^*$) 
the resulting curves of runs~$1$ and $2$ almost overlap and the amplitude 
of oscillations is nearly equal to unity, because the bed shear stress approaches that of a 
Stokes boundary layer in both runs. %
%
%Similar conclusions would be drawn by normalising the bed shear stress with 
Consequently, the quantity $\frac{1}{2}\densf^*\U^*\om^*\del^*$ %, which 
is a relevant scale for the bed shear stress. %
Hence, for a given value of $\mob$, by increasing the value of $\Rdel^*$ 
the Shields parameter decreases (like in the present case) until turbulence 
appears and further modes of sediment transport occur (e.g. saltation). %
Moreover, it can be noted that the maximum value 
$\max\left(\shields_{(St)}\right) = 
\sqrt{2}\frac{\mob}{\Rdel}$ 
equals the maximum value of the Sleath number, i.e. approximately the maximum effect of 
%the difference of pressure between the front and back sides of 
the imposed pressure gradient on %
an isolated particle, 
$\max\left(\mathcal{S}\right) = 
\frac{\mob}{\Kc} = 
2\frac{\ds^*}{\del^*}\frac{\mob}{\Rdel}$, if $d^*/\del^*\sim 0.7$. %
Therefore, even though the maximum shear stress and the maximum imposed pressure gradient 
are reached at different phases of the oscillation period, in the present cases 
\citep[and in most of the experiments made by][]{blondeaux1988} viscous effects prevail. %
It can be useful to point out that typically, in the field, the ratio $\ds^*/\del^*$ does not significantly vary with respect to the other parameters (e.g. for $0.2$~mm$<\ds^*<1$~mm and $\TT\sim 10$~s, $0.1<\ds^*/\del^*<0.6$), thus the ratio $\frac{\mob}{\Rdel}$ 
can be practically used as the only parameter driving the sediment flow rate 
(as long as the flow is not turbulent). %
It can be inferred from the present results that the growth rate of ripples is related to the maximum sediment flow rate. 
In particular, the growth rate of ripples increases if the maximum bed shear 
stress is not much larger than the critical value of incipient sediment motion, 
sediments being more sensitive to the effect of the steady streaming. %
Consequently, if the ratio $\frac{\mob}{\Rdel}$ is close to 
$\shields_{cr}/\sqrt{2}\sim 0.035$ ripples form more rapidly. %
In fact, for runs~$1$ and $2$, $\frac{\mob}{\Rdel}$ is equal to $0.076$ and $0.110$, respectively, and ripples form much more slowly in case~$2$. %
The relevance of $\frac{\mob}{\Rdel}$ for the prediction of ripple genesis was also emphasized by \citet{blondeaux1990} (see figure~$11$) because it is related to the ratio $\frac{\sav{\pse^*}_s}{\wln^*}$. %
In particular, the dimensionless parameter used by \citet{blondeaux1990} and by other authors before was $\frac{d^*}{(s-1)g^*\TT^{*2}}$ which is equal to %$\frac{1}{4\pi^2}\frac{\mob}{\Kc^2}$ 
$\frac{1}{\pi^2}\frac{\mob}{\Rdel^2}\frac{d^{*2}}{\delta^{*2}}$ and was empirically 
found controlling the ripple wavelength. %
%\mun{Aggiungere nota in cui si indica il parametro richiamato da \citet{blondeaux1990}.}
}%

\section{Conclusions}\label{conc}

%    Conclusions
The origin and development of ripples in an oscillatory flow was investigated 
by means of direct numerical simulations. %
Two experiments were reproduced which were carried out by using %
\mmr{a}{} %
medium sand 
at moderate values of the Reynolds number. %
The experiments significantly differed in the frequency and amplitude 
of the free-stream velocity oscillations (i.e. both in the Stokes and particle 
Reynolds numbers, $\Rdel$ and $\Rd$). %
After approximately ten oscillations, two-dimensional patterns arose which 
then coarsened turning into rolling-grain ripples. %
The wavelengths characterizing the ripples in the simulations, in the limits set 
by the domain size, are comparable with those observed in the experiments and with 
the predictions obtained by linear stability analysis. %
The bed surface is identified for each discrete instant. %
The Fourier analysis of the bed profile shows that, after an initial transient where
patterns form then merge or disappear, a few wave numbers grow in amplitude and finally 
one wave number becomes dominant. %
Ripples form clearly in one of the two simulations (run~$1$) while two-dimensional 
patterns are observed in the second simulation (run~$2$), since the dynamics of the bed %
\mmr{is}{are} %
somewhat slower in the latter case. %
In %
\mmr{the former case}{run~$1$} %
the growth of the bed-surface 
fluctuation amplitude normalised by 
the particle diameter is found to follow 
an exponential trend %
\mmr{of}{with} %
exponent equal to $1.46\cdot 10^{-2}~\om^*t^*$. %
The secondary flow arising from the flow instability consists %
\mmr{in}{of} %
steady recirculating cells 
which are responsible for the formation of ripples, since they tend to pile up the sediment 
particles at the nodes where streamlines converge and to scour where streamlines diverge 
close to the bed surface. %
\mmt{%
The evolution of ripples and the development of recirculating cells are strictly related. %
}%
Ripples of run~$1$ exhibit an asymmetric shape for most of the oscillation period, 
with the lee side steeper than the stoss side, except in the phases characterised 
by the largest bed shear stress when the %
\mmr{crest of ripples}{ripple crests} %
migrates in the direction 
of the mean flow. %
The sediment particles at the flow-bed interface (top-layer particles) are tracked 
during the wave cycles and the velocity and the drag force are computed. %
Two distinct kinds of particle motion are identified: most of \mmt{the} top-layer particles, 
in particular those lying in the troughs of ripples, roll for $\mathcal{O}(1)~d^*$ in the 
flow direction then they stop. %
The excursion of these particles, i.e. the displacement in the streamwise direction 
that they experience for each half cycle, is found to scale with the wavelength of 
ripples 
%mm
%(i.e. the dominant wavelength, $\wln^*$) 
for the present simulations. %
Similarly, the drag force and, more weakly, the velocity of these ``slow'' particles 
scale with reference quantities obtained as combinations of $\om^*$, $U_0^*$ and $\delta^*$. %
\mmr{Instead}{However}, the sediment particles lying on the crest of ripples (crest particles) are 
subjected to stronger drag force which causes large excursions in some cases of %order 
$\mathcal{O}(0.1)~\fse^*$, i.e. comparable with the fluid excursion far from the bed. %
\mmt{Therefore, such particles are provided with larger momentum than others. }%
These ``fast'' particles, though they do not saltate, encounter several collisions 
with other particles that contribute to increase the variance of quantities 
associated with their motion. %
It is found for the present cases that the wider %
\mmr{is}{} %
the difference of motion between 
``slow'' and ``fast'' particles, the more rapid %
\mmr{is}{} %
the growth of bedforms \mmt{is}. %
%
%\mmr{}{%
%
\mmc{Indeed, some particles are provided with larger momentum than others. }%
In this sense the origin of ripples can be seen in analogy with 
the phenomenon of segregation of sediments of different size or density, 
since in both cases %in the present case 
a non-uniform distribution of momentum is transferred from the flow to the sediments, 
in one case because of 
the non-uniform distribution of the mass of sediment grains, 
while in the present case because of the non-uniform distribution of the velocity field 
(due to the presence of the recirculation cells). %
%}% 
%

%
Finally, the sediment flow rate is computed and compared with global quantities characterising 
the fluid-sediment interaction. %
The Shields number and the dimensionless (external) pressure gradient are considered. %
A fair correlation between the sediment flow rate and the Shields number is found in the phases 
of the oscillation period when the bed shear stress reaches the maximum value. %
In such phases the predictions obtained by means of the Meyer-Peter \& M\"uller formula, i.e.  
based on steady flow regime and uniquely on the value of the Shields number, are approached. %
However, in the phases of the oscillation period characterised by small values of the bed 
shear stress and large values of the (external) pressure gradient, a significant sediment 
flow rate was observed which cannot be explained on the basis of the instantaneous Shields 
number, which is actually vanishing. %
Thus, the Shields number should be combined with the dimensionless pressure gradient to improve 
the accuracy of prediction of the sediment flow rate. %
\mmr{Concluding}{In conclusion}, for the purpose of modelling the formation of bedforms under sea waves, 
the effect of the unsteadiness on the transport of sediments is remarkable %, even 
in the absence of turbulent events as in the present cases. %
\mmt{%
For the range of values of $d^*/\delta^*$ that typically characterise a sandy seafloor, the parameter $\mob/\Rdel$ controls the growth rate of ripples. %
In particular, the closer $\mob/\Rdel$ is to $\shields_{cr}/\sqrt{2}$, the more rapid the formation of ripples results. %because the effect of recirculating cells on the mean particle dynamics is more pronounced. %
}%

An extension of the present investigation aimed at exploring the regions of the parameter space 
characterised by the turbulent flow regime would be of immense help for the development of 
reliable sediment transport models and for the estimation of the bed evolution. %
\mmt{%
Considering the high computational cost and the formidable simulation %
%MU
%(elapsed) 
time required by the present simulations, which exceeded the $10$~million CPU hours and approximately $1$~million time steps (i.e. running for $\sim480$ days on $64$ ``Ivy Bridge'' computing nodes), the direct numerical simulation of the formation of bedforms in a turbulent oscillatory flow is not yet feasible as it would require much larger domain and finer spatial and temporal resolutions than the present ones. %
Nonetheless, fundamental insights on the mechanics of sediment transport in a turbulent oscillatory boundary layer could be obtained by reducing the size of the computational domain to that required by turbulence to develop, namely to the minimal flow unit. %
}%

\smallskip
%%    Acknowledgement
This study has been funded by the Office of Naval Research (U.S., under the research project no.~1000006450 - award no.~N62909-17-1-2144) and by the Deutsche Forschungsgemeinschaft (Germany, project UH 242/4-2). %
We acknowledge the generous support from CINECA (Bologna, Italy) for the computational resources provided on FERMI under grant TEST SEA (PRACE 7th-call) and from the Steinbuch Center for Computing (KIT, Karlsruhe) for the resources provided on ForHLR I under the grant DNSBESTSEA. %
The Authors wish to thank Paolo Blondeaux and Giovanna Vittori for the fruitful discussions in particular concerning the interpretation of the bed evolution and the description of the experiments they carried out in Cambridge in 1988. %
The numerical investigations were performed during the time the second
author (AGK) was employed by the Institute for Hydromechanics, Karlsruhe
Institute of Technology (KIT). %
%

%\appendix

%\include{nomenclature}
%\printnomenclature


\bibliographystyle{elsarticle-harv} 
\bibliography{\refdir Refbib.bib}
\ifnotes
\section*{Temporary notes}

\input{notes.tex}
\input{old_figures.tex}
\fi % \ifnotes
\end{document}